\documentclass{aa}
\usepackage{graphicx}
\usepackage{txfonts}
\usepackage{natbib}
\bibpunct{(}{)}{;}{a}{}{,}
\usepackage{lscape}
\usepackage{rotating}
\usepackage{amsmath}
\usepackage[deletedcolor=red, addedcolor=blue]{changes}
\setdeletedmarkup{\textcolor{red}{\sout{#1}}} %
\usepackage[colorlinks,linkcolor=blue,anchorcolor=blue,citecolor=blue,]{hyperref}

\newcommand{\hi}{\mbox{H\,\textsc{i}}}

\begin{document}

\title{Accurate polarization calibration of FAST spectral data for measurements of Zeeman splittings of OH megamasers in IRAS\,02524+2046}
\author{L. G. Hou\inst{1,2,3}, X. Y. Gao\inst{1,2,3}, Tao Hong\inst{1,2,3}, J. L. Han\inst{1,2,3}}

\offprints{L.G. Hou. \email{lghou@nao.cas.cn}}

\institute{National Astronomical Observatories, CAS, Jia-20 Datun Road, Chaoyang District, Beijing 100101, PR China \and  School of Astronomy and Space Sciences, University of Chinese Academy of Sciences,
     Beijing 100049, China \and State Key Laboratory of Radio Astronomy and Technology, Beijing 100101, China}

\date{Received; accepted}

\abstract 
{An accurate polarization calibration is essential for a spectral data analysis and Zeeman splitting measurements. Two anomalies challenge our understanding of OH megamasers in {\it IRAS}\,02524+2046: an unexplained 1667/1665 MHz flux-ratio deviation, and complex Stokes $V$ signatures. Well-calibrated sensitive polarization observations are required to understand them.}
{We develop a polarization calibration solution for the $L$-band 19-beam receiver installed on the Five-hundred-meter aperture spherical radio telescope (FAST) to achieve a high calibration accuracy and thus enable accurate measurements of the OH megamaser properties in {\it IRAS}\,02524+2046.}
{We determined the Mueller matrix solution for spectral observations across the 1050$-$1450~MHz frequency range with an accuracy of $\sim0.01\%-0.08\%$ for circular polarization. 
We then applied it to FAST observational data of {\it IRAS}\,02524+2046.}
{Our results show narrower emission line components in the OH megamasers than previously reported, which are indistinguishable in the total power spectrum, but are detected in the circular polarization spectrum.
The 1667~MHz OH megamaser emissions probably span a wide velocity range from $v_{\rm helio}$~$\sim$~54\,750 to $\sim$~53\,580~km~s$^{-1}$, indicating greater complexity than previously recognized.
Our fit of the total power and circular polarization spectra for {\it IRAS}\,02524+2046 revealed ten line components with significant Zeeman splitting ($>3\sigma$), indicating in situ magnetic fields with a strength of approximately $-$24.5~mG to +20.6~mG, most of which (8/10) have positive values.}
{}



\titlerunning{Polarization calibration and OH Megamasers in {\it IRAS}\,02524+2046}
\authorrunning{L. G. Hou et al.}

\maketitle
\section{Introduction}
\label{sect:intro}
An accurate polarization calibration is essential for analyzing spectral radio data. In general, the Mueller matrix for the entire observation system is determined through multiple observations of a standard polarization calibrator (e.g., 3C\,286) at various parallactic angles \citep[][]{heiles01,robi08,rh21,clh22,chl+24}. These polarization observations for an accurate calibration require significant telescope time. 

The Five-hundred-meter aperture spherical radio telescope (FAST) is currently the world's largest single-dish radio telescope \citep{nan+2006}. With its 300 m illuminated aperture and the sensitive $L$-band 19-beam receiver covering 1000$-$1500~MHz \citep{fast20}, it possesses exceptional capabilities for observing pulsars \citep[e.g.,][]{gpps,zhx+23,xhw+22}, fast radio bursts \citep[e.g.,][]{zhj+23}, radio continuum emission \citep[e.g.,][]{gao+22}, and spectral lines \citep[e.g.,][]{hong+22,hou23}. Since its commissioning in 2019, FAST has conducted multiple observations of polarization calibrators. 
A circular polarization calibration accuracy on the order of $10^{-4}$ for the central beam has been achieved by \citet{clh22}. Based on observations of the calibrators 3C\,286, 3C\,48, and 3C\,138, \citet{chl+24} recently characterized the temporal variations of the Mueller matrix elements for the polarization calibration for the 19 beams. An accurate calibration for all beams requires regular observations of polarization calibrators. The currently available observation data from FAST are insufficient for the purpose.

During spectral line observations, the calibration signals are routinely injected into the system in on-off mode for a rapid 
calibration. This approach enables us to determine gain and phase differences and variations between the receiver's two linearly polarized channels, as demonstrated in continuum polarization studies \citep[e.g.,][]{gao+22,xzs+23} and pulsar polarization research \citep[e.g.,][]{whx+23,wwh+24}. 
Since the commissioning of early scientific observations in 2019, FAST has accumulated a substantial volume of spectral line data using its
$L$-band 19-beam receiver. 
The spectral backends record all four polarization products ($XX$, $YY$, Re[$X^{*}Y$], and Im[$X^{*}Y$]) from the radio signals received through the two orthogonal linear polarizations $X$ and $Y$.

We develop the procedures for accurately calibrating the full-polarization spectral data obtained by FAST. 
We then apply our approach to observations of OH megamasers in {\it IRAS}\,02524+2046, which were observed in our projects to validate the data processing pipeline and Stokes $V$ sign convention.
The unprecedented sensitivity of FAST reveals new properties of OH megamasers in this galaxy. We briefly introduce the {\it IRAS}\,02524+2046 and FAST observations in Section~\ref{sect:data}. The polarization calibration procedures are presented in Section~\ref{sect:cal}. The new results for OH megamasers in {\it IRAS} 02524+2046 are presented in Section~\ref{sect:results}. Our discussions and conclusions are given in Section~\ref{sect:con}.

\section{OH megamasers of {\it IRAS}\,02524+2046 and FAST observations}
\label{sect:data}

The galaxy {\it IRAS}\,02524+2046 is a starburst galaxy \citep[][]{pwz+20} and hosts luminous OH megamasers exhibiting unusual line ratios \citep[][]{dg02,mhe13b}. The OH 
spectrum of this galaxy presents broad-line components together with multiple strong and narrow components of the 1667 and 1665~MHz transitions. Some spectral components show day-to-day variation \citep{darl05,darl07}. No satellite lines of the OH ground state near the rest frequency of 1612~MHz or 1720~MHz were detected \citep[][]{mhe13b}. Observations with a high spatial resolution reveal that the compact maser emission line sources are distributed across a region of $\sim$210~$\times$~90~pc \citep[][]{pwz+20,wsz+23}.

There are two unresolved issues regarding the OH megamasers in {\it IRAS}\,02524+2046. One issue is the anomalous flux ratio for the pair of the 1667 and 1665~MHz lines \citep[][]{dg02,mhe13b}, and the other issue is the complex features in the Stokes $V$ spectrum \citep[][]{mh13}. The high-velocity feature at $v_{\rm helio} \sim54\,725$~km~s$^{-1}$, attributed to the OH 1665~MHz transition in previous studies \citep[][]{dg02,mhe13b,pwz+20}, shows an unusually strong integrated flux density compared to the 1667~MHz line at a heliocentric velocity $v_{\rm helio} \sim54\,300$~km~s$^{-1}$. 
Unless otherwise specified, the $v_{\rm helio}$ of the observed spectrum throughout this work was calculated using the rest frequency ${\nu}_0=$~1667.3590~MHz.
The line ratio $R_H=F_{1667}/F_{1665}$, where $F_{\nu}$ represents the integrated flux density across each emission line, is about 1.3 \citep{mhe13b} or 1.4 \citep{dg02}, which is in the range of thermal emission values of $1.0 - 1.8$. It presents a challenge to the pumping model of OH megamasers \citep[][]{le08}, which predicts that the 1667~MHz line dominates the 1665~MHz line for line widths exceeding $\sim$2~km~s$^{-1}$, as commonly seen in OH megamasers. Even after multi-Gaussian decomposition, both lines of the pair of 1667 and 1665~MHz transitions maintain significantly broader line widths ($>$ 2~km~s$^{-1}$). On the other hand, it is difficult to fit the Stokes $V$ spectrum features \citep[][]{mh13} to demonstrate the Zeeman splitting and derive the magnetic fields of OH megamasers, because significant residuals with prominent peaks and dips persist in all fitting attempts.

{\it IRAS}\,02524+2046 was observed by FAST during two observation sessions (see below) using the central beam of the $L$-band 19-beam receiver with a zenith angle from approximately 24.7$^\circ$ to 9.4$^\circ$ and a small system temperature change in a range of about $T_{\rm sys} \sim 23$~K to 20~K \citep{fast20}.
The two linear polarization signals $X$ and $Y$ were extracted by an orthomode transducer for each of the 19 beams, making this system particularly well suited for precise measurements of circular polarization signals \citep[][]{heiles01}. 

\begin{table*}[!t]
\caption{OH megamaser observation parameters for {\it IRAS} 02524+2046.} 
\label{tabsource}    
\tabcolsep 16pt 
\centering                             
\begin{tabular*}{\textwidth}{lc}
\hline\hline
Observation date          &  11 and 12, August 2023 \\
Observed frequency range  & 1000 $-$ 1500 MHz \\  
Effective frequency range & 1050 $-$ 1450 MHz \\
Beam size                 & $\sim$2.8$^\prime-2.9^\prime$@1420~MHz \\
Polarization products     & $XX$, $YY$, Re[$X^{*}Y$], Im[$X^{*}Y$]) \\
Channel number            & 1024 k \\
Sampling time             & 0.5~s \\
Integration time          & 110~minutes \\
\hline                         
Redshift of {\it IRAS} 02524+2046 & 0.1814 $\pm$ 0.0002$^\dagger$  \\
Rest frequencies of the four OH lines$^\ddagger$  & 1612.2310, 1665.4018, 1667.3590, 1720.5300 MHz \\
Observed frequencies$\ast$          & 1364.6783, 1409.6850, 1411.3416, 1456.3484 MHz \\
\hline
\end{tabular*}
\tablefoot{$^\dagger$ Redshift derived from optical spectrophotometry \citep{dg06}. \\
$^\ddagger$ Rest frequencies of the four OH transitions were taken from https://pml.nist.gov/cgi-bin/micro/table5/start.pl\\
 $\ast$ Observed frequencies were calculated using their rest frequencies and the galaxy redshift.}
\end{table*}

The first observation was carried out on 11 August 2023 using the central beam of the $L$-band 19-beam receiver. The TrackingWithAngle mode was adopted to track the target for approximately one hour, with a pointing accuracy of about $8''$ \citep{fast20}, a main-beam efficiency of $\sim$0.63, and a half-power beam width of $\sim$2.8$^\prime$~\citep[][]{fast20} or $\sim$2.9$^\prime$~\citep[][]{ccl+25} at 1420~MHz. During the observations, the receiver continuously rotates to compensate for the real-time field rotation. Additionally, the feed angle can be preset to any value between $-80^\circ$ and $+80^\circ$. We maintained the default 0$^\circ$ setting of the feed angle for these observations. A calibration signal from a low-noise diode with an amplitude of about 1.1~K \citep[][]{fast20} was periodically injected for 1~s every 16~s, which can be used to calibrate the temperature scale and the polarization performance of the system \citep[][]{smg+21}. The polarization signals were recorded across 1024~k channels covering the observing frequencies from 1000~MHz to 1500~MHz, corresponding to a frequency resolution of 0.477~kHz (equivalent to 0.10~km~s$^{-1}$ velocity resolution at the 1420~MHz $\hi$ line).

The second observation was conducted on 12 August 2023 using identical instrument settings, but with the feed angle fixed at 45$^\circ$ after compensating for the field rotation to examine potential polarization side-lobe effects. We note that {\it IRAS}\,02524+2046 is located at Galactic coordinates ($l, b$) = ($158.0^\circ$, $-33.3^\circ$). The spectral line data of the foreground Galactic $\hi$ are recorded simultaneously, which provides an independent assessment of the polarization calibration quality, as discussed in subsequent sections.

For the polarization calibration, we obtained five drift-scan observations of the standard calibrator 3C\,286 on 19 August 2023 at different rotation angles of $-60^\circ$, $-30^\circ$, $0^\circ$, $+30^\circ$, and $+60^\circ$~\citep[e.g.,][]{fast20,clh22,chl+24}. Given the strong flux density of 3C\,286, we employed a high noise diode with an amplitude of about 12.5~K \citep[][]{fast20} that injected calibration signals in a period of 1~s during the drift scans, with spectral data recorded every 0.5~s. Additionally, we incorporated FAST observations of 3C\,286 on 7 and 14 August 2023 using the multi-beam calibration mode during the scheduled maintenance periods to calibrate the flux density.

\begin{figure}
  \centering
    \includegraphics[width=0.44\textwidth]{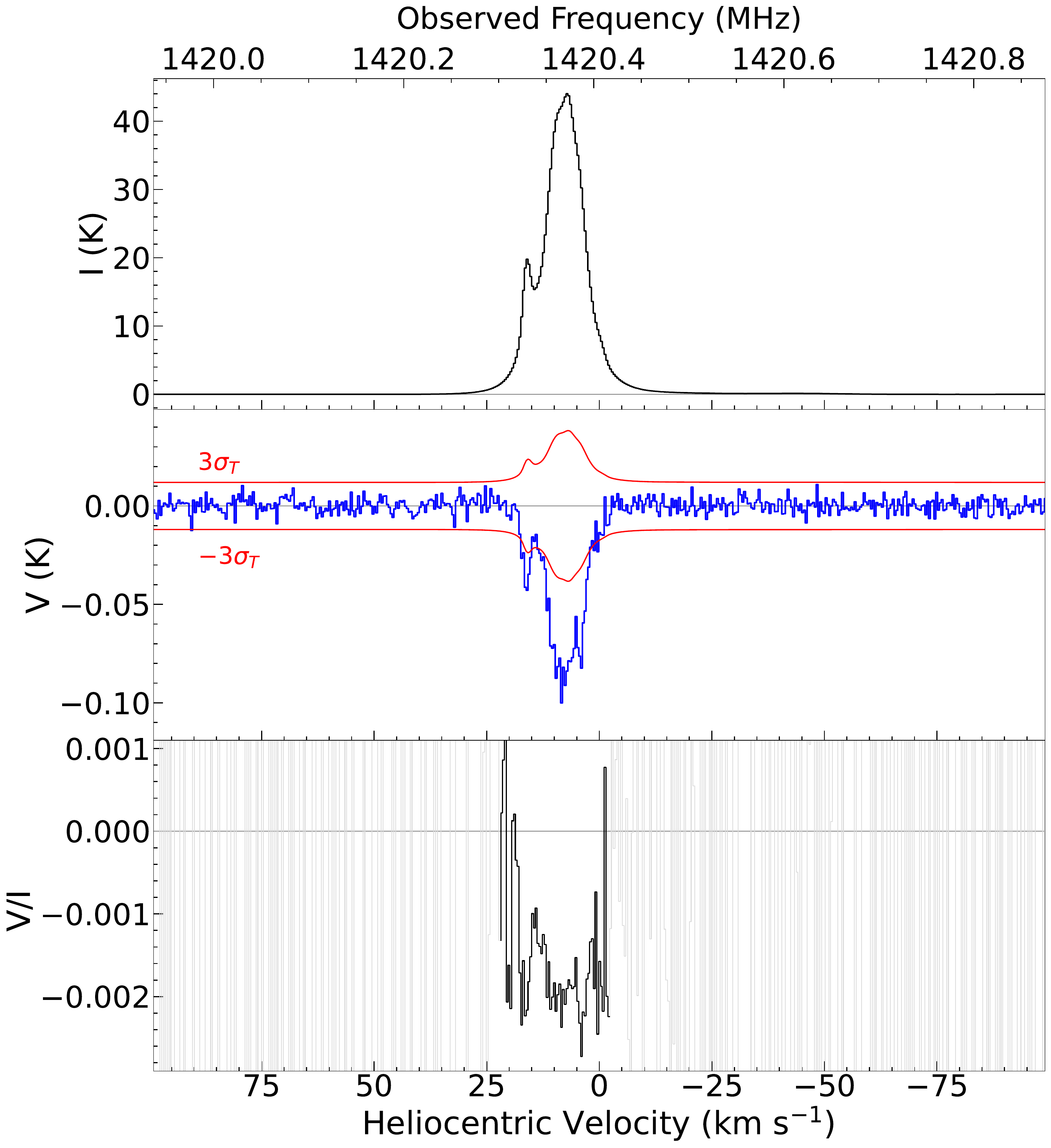} \\
 \vspace{0.25cm}
  \includegraphics[width=0.44\textwidth]{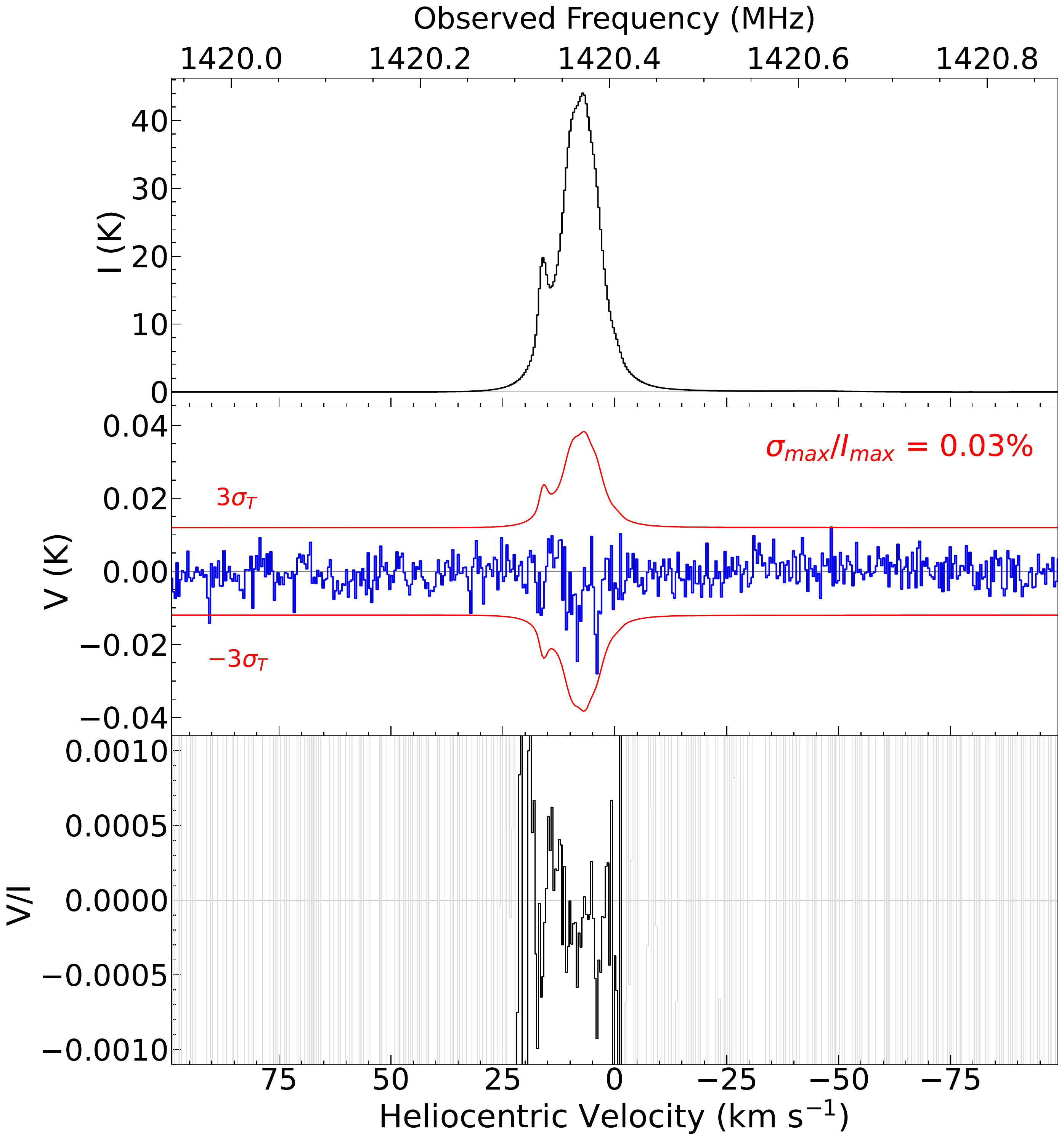} 
    \caption{Calibrated results of the Galactic $\hi$~21~cm lines using the two approaches described in Sect.~\ref{sect:noise} and Sect.~\ref{sect:mul}. These FAST observations targeted {\it IRAS}\,02524+2046 at Galactic coordinates ($l,b$) = (158.0$^\circ$, $-$33.3$^\circ$), with a total integration time of 110 minutes. The {\it upper panels} demonstrate that the quick calibration method using injected reference signals achieves polarization measurement accuracy of $\sim$0.2\%.  The $\sigma_T$ lines are derived by considering the contributions from $\hi$ emission features following the method of \citet{jhh+23}. The {\it lower panels} show results after a full Mueller matrix solution and leakage correction, reaching a higher calibration accuracy of $\sim$0.03\%.
   }
   \label{galhi}
\end{figure}

\section{Calibration of polarized spectral data}
\label{sect:cal}

Based on the intensities of periodically injected reference signals, the four polarization products ($XX$, $YY$, Re[$X^{*}Y$], and Im[$X^{*}Y$]) were first calibrated to the scale of the antenna temperature $T_A$ and were then converted into the heliocentric frame by correcting for the Doppler effect. 
The radio frequency interference (RFI) in the data was inspected and removed as done by \citet[][]{hou23}. Afterward, the observed Stokes parameters were derived from the four polarization products using
\begin{equation}
\begin{bmatrix}
I_{\rm obs} \\
Q_{\rm obs} \\
U_{\rm obs} \\
V_{\rm obs} \\
\end{bmatrix}
  = \begin{bmatrix}
XX + YY \\
XX - YY \\
2Re[X^{*}Y] \\
\pm 2Im[X^{*}Y] \\
  \end{bmatrix},
 \end{equation}
where the sign ambiguities in the polarization signals caused by internal cable connections were solved by comparing our results with published values. Following \citet{rqh08} and \citet{mh13}, we adopted the IAU definitions of Stokes $V = RCP - LCP$, in accordance with the IEEE standard that the right circular polarization ($RCP$) rotates clockwise as viewed from the radio source, and the left circular polarization ($LCP$) rotates counterclockwise. 

The observed Stokes parameters ($I_{\rm obs}$, $Q_{\rm obs}$, $U_{\rm obs}$, and $V_{\rm obs}$) are the products of the radio source polarization signals modified by the receiving system of a telescope, as 
\begin{equation}
\begin{bmatrix}
I_{\rm obs} \\
Q_{\rm obs} \\
U_{\rm obs} \\
V_{\rm obs} \\
\end{bmatrix}
= M_{tot} \bullet 
\begin{bmatrix}
I_{\rm source}' \\
Q_{\rm source}' \\
U_{\rm source}' \\
V_{\rm source}' \\
\end{bmatrix}.
\end{equation}
Here, $M_{tot}$ is the polarization transfer function of the receiving system \citep[][]{heiles01}, which we discuss below. The polarized source signals are modified by 
the parallactic angle, $\theta$, via  
\begin{equation}
\begin{bmatrix}
I_{\rm source}' \\
Q_{\rm source}' \\
U_{\rm source}' \\
V_{\rm source}' \\
\end{bmatrix}
= 
  \begin{bmatrix}
1   &  0 & 0 & 0\\
0   & cos2\theta & sin2\theta & 0 \\
0   & -sin2\theta & cos2\theta & 0 \\
0   &   0     &   0    & 1 \\
  \end{bmatrix} \bullet 
\begin{bmatrix}
I_{\rm source} \\
Q_{\rm source} \\
U_{\rm source} \\
V_{\rm source} \\
\end{bmatrix}.
\end{equation}
For FAST observations, the feed is always rotated to compensate for field rotation, so this modification is usually not present as $\theta$ was set to 0$^\circ$. To derive intrinsic source properties from the observed Stokes parameters, the polarization transfer function of the receiving system \citep[][]{heiles01},  $M_{tot}$, has to be solved, which can be expressed as 
\begin{equation*}
\setlength\arraycolsep{5pt}
  \begin{bmatrix}
1   &  X_{I2Q}  & X_{I2U} &  2\epsilon\,sin\phi\\
{\Delta G}/{2} & X_{Q2Q} & X_{Q2U} & {\Delta G}\epsilon\,sin\phi \\
2\epsilon\,cos(\phi + \psi) & -sin2\alpha\,cos\psi & cos2\alpha\,cos\psi & -sin\psi\\
2\epsilon\,sin(\phi + \psi) & -sin2\alpha\,sin\psi & cos2\alpha\,sin\psi & cos\psi \\
  \end{bmatrix}
\end{equation*}
for the dual linear feeds of the FAST $L$-band 19-beam receiver \citep{chl+24}. Here, $\epsilon$ denotes the imperfection of the feed in producing nonorthogonal
polarizations, $\phi$ is the phase angle at which the voltage coupling $\epsilon$ occurs, $\alpha$ measures the voltage ratio of the polarization ellipse produced when observing a pure linear polarization signal, $\Delta G$ indicates the error in the relative intensity calibration of the two polarization channels, and $\psi$ is the phase difference between the reference noise signal and the incoming radiation from the sky, $X_{I2Q} = -2\epsilon\,cos\phi\,sin2\alpha +\frac{\Delta G}{2}\,cos2\alpha$,  $X_{I2U} = 2\epsilon\,cos\phi\,cos2\alpha + \frac{\Delta G}{2}\,sin2\alpha$, $X_{Q2Q} = -\Delta G\,\epsilon\,cos\phi\,sin2\alpha + cos2\alpha$, and $X_{Q2U} = \Delta G\,\epsilon\,cos\phi\,cos2\alpha + sin2\alpha$.
To determine the intrinsic source properties from the observed Stokes parameters, we investigated two different calibration approaches.

\subsection{Polarization calibration resting on injected reference signals}
\label{sect:noise}

Polarization calibration using the injected reference signals has been successfully applied to various FAST observations, including studies of radio continuum sources and pulsars \citep[e.g.,][]{smg+21,whx+23}. Instead of determining all Mueller matrix elements \citep[e.g.,][]{heiles01,robi08,rh21}, this simplified approach focuses on solving for the two dominant parameters, $\Delta G$ and $\psi$, which account for the mismatch between the amplitudes and phases of the gains of the two orthogonal linear feeds, respectively. This method offers two key advantages. One advantage is that no additional observations of the polarization calibrator are needed, which saves time. The other advantage is that the time-dependent small variation in the system polarization characteristics, even during the observations, can be monitored and corrected.

Following the method of \citet{smg+21}, we determined $\Delta G$ and $\psi$
for each frequency channel and applied these corrections to the five 3C\,286 drift scans obtained at different rotation angles. We measured a linear polarization fraction $P_{src} = \sqrt{Q^2+U^2}/I = 9.3\% \pm 0.5\%$ and polarization angle $\chi = \frac{1}{2} \arctan \frac{U}{Q} = 25.5^\circ\pm0.8^\circ$ at 1400~MHz after ionospheric correction using the ionFR package \citep[][]{ionfr}. These results are roughly consistent with reference values of $P_{src} = 9.90\% \pm 0.003\%$ \citep{tl24} and $\chi \sim 27.8^\circ - 28.6^\circ$ at 1400~MHz \citep{tl24} and with our results in Sect.~\ref{sect:mul}. According to \citet{chl+24}, the calibrated polarization percentages and polarization angles of 3C\,286 measured by FAST at 1420\,MHz show $P_{src}$ values ranging from $\sim$9.3\% to 9.8\% during 2019 to early 2023, with $\chi$ values varying between $\sim 28^\circ$ and 32$^\circ$. However, the calibrated $V/I$ values deviate from zero across 1050$-$1450~MHz, ranging from about $-0.5\%$ to $+0.4\%$. The absolute median and mean values are $0.15\%\pm0.01\%$ and $0.17\%\pm0.01\%$, respectively, indicating that the calibrated $V$ by this method is dominated by the uncompensated leakage of $I$ at the $0.2\%$ level.

To assess the reliability of this simple calibration method, we examined the calibrated spectra of {\it IRAS}\,02524+2046 around the Galactic $\hi$~21~cm line region. For this target direction, circular polarization of the Galactic $\hi$~21~cm line is expected to be undetectable given our observational integration time. The upper panels of Fig.~\ref{galhi} show that the calibrated $V$ spectrum has a very similar shape as the $I$ profiles near the heliocentric velocity of $-10$~km~s$^{-1}$, indicating that $V$ has been dominated by the uncompensated leakage of $I$ into $V$ at approximately $-0.2\%$ near 1420.4~MHz. For most of the spectra data accumulated by FAST in the past few years, when observational data for determining the full Mueller matrix are lacking and the simple calibration method based on the injected reference signals is applied, the uncompensated leakage of $I$ into $V$ at the $\sim$0.2\% level cannot be corrected.

\subsection{Polarization calibration based on observations of 3C\,286}
\label{sect:mul}

\begin{figure}
  \centering
  \includegraphics[width=0.47\textwidth]{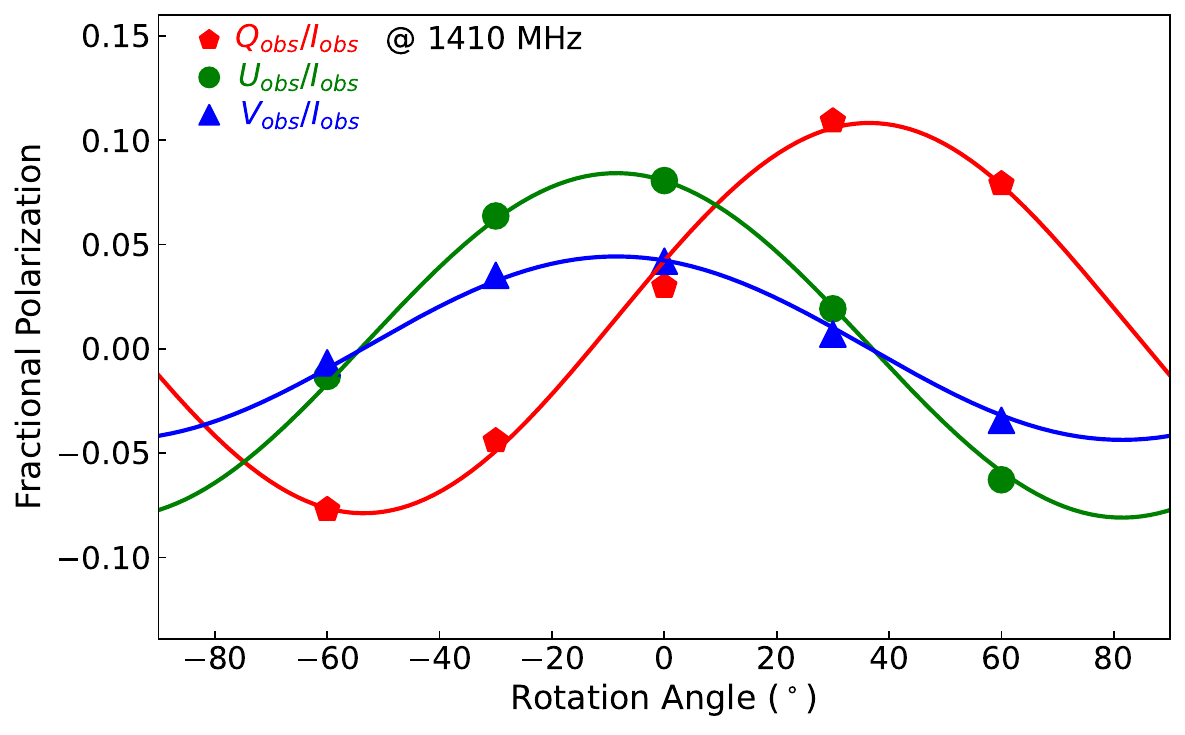} \\
  \includegraphics[width=0.47\textwidth]{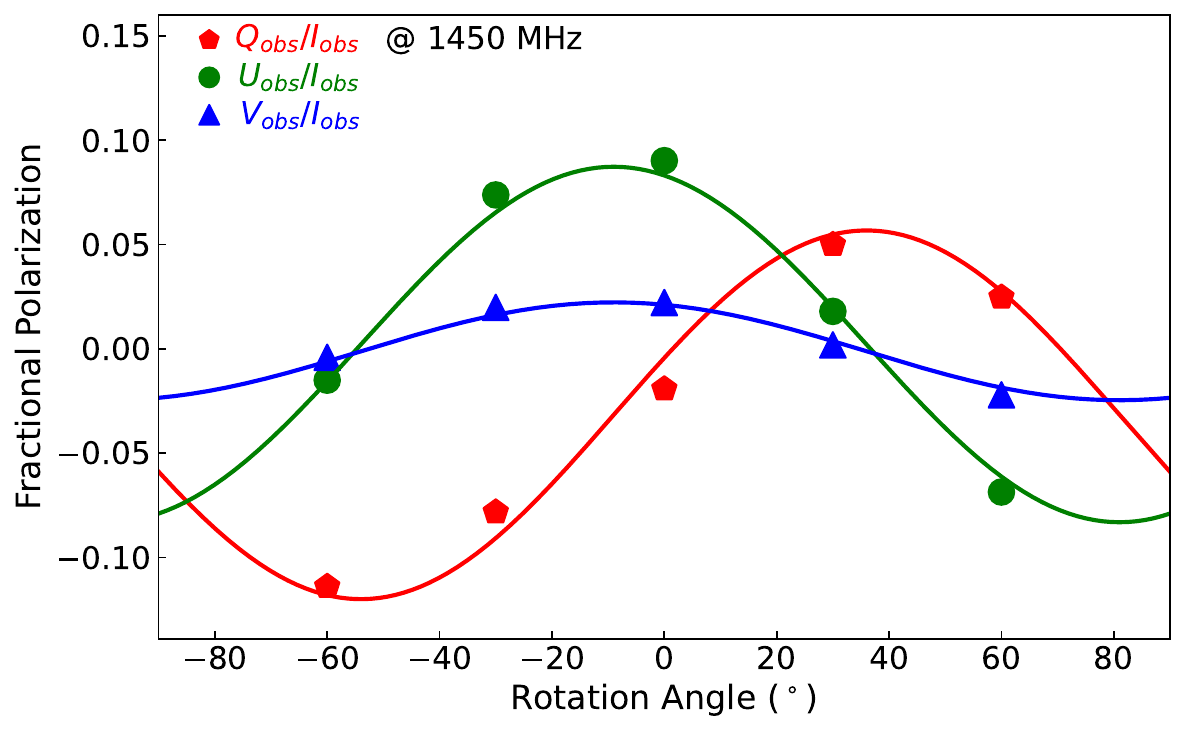}
   \caption{Fitting the observed fractional polarizations in two example frequency channels of 1410~MHz ({\it upper}) and 1450~MHz ({\it lower}) to derive the Mueller matrix elements. The different symbols indicate different observed fractional polarization components, as shown in the plots. The solid curves show the fitting results. The data were obtained from FAST observations on 19 August 2023.}
   \label{muller}
\end{figure}

The standard polarization calibrator 3C\,286 was used to determine the Mueller matrix coefficients for the FAST telescope system. In general, the intrinsic circular polarization $V$ of 3C\,286 is assumed to be zero at $L$-band, as confirmed by recent observations \citep[e.g.,][]{tl24}.
To derive the Mueller matrix from FAST's five 3C\,286 drift scans, we implemented the method of \citet{heiles01}, and we refer to the documentation for the package Robishaw/Heiles SToKes (RHSTK)\footnote{https://w.astro.berkeley.edu/\%7Eheiles} as well as the procedures described in \citet{chl+24}. 
We fit the observed fractional polarizations ($Q_{obs}/I_{obs}$, $U_{obs}/I_{obs}$, and $V_{obs}/I_{obs}$) for each frequency channel. The fitting results for two frequency channels are shown in Fig.~\ref{muller}. 
Across the frequency ranges of 1050$-$1150~MHz and 1300$-$1450~MHz, the fit parameters vary as follows: $\Delta G$, $-3.0$\% to 2.9\%; $\psi$, $-0.8^\circ$ to $1.4^\circ$; $\epsilon$, $-0.38$\% to $0.29$\%; $\phi$, $16^\circ$ to $177^\circ$; and $\alpha$, $-0.3^\circ$ to $-0.2^\circ$ (see Fig.~\ref{pamuller}). The Mueller matrix parameters for the 1150$-$1300~MHz range were not determined due to severe RFI.
The derived Mueller matrix was then applied to observational data to obtain corrected Stokes parameters ($I$, $Q$, $U$, $V$). The sign of Stokes $V$ was verified using previously known results of OH megamasers from {\it IRAS}\,02524+2046 \citep[][]{mh13}.                            

\begin{figure*}[t]
\centering
\includegraphics[width=0.85\textwidth]{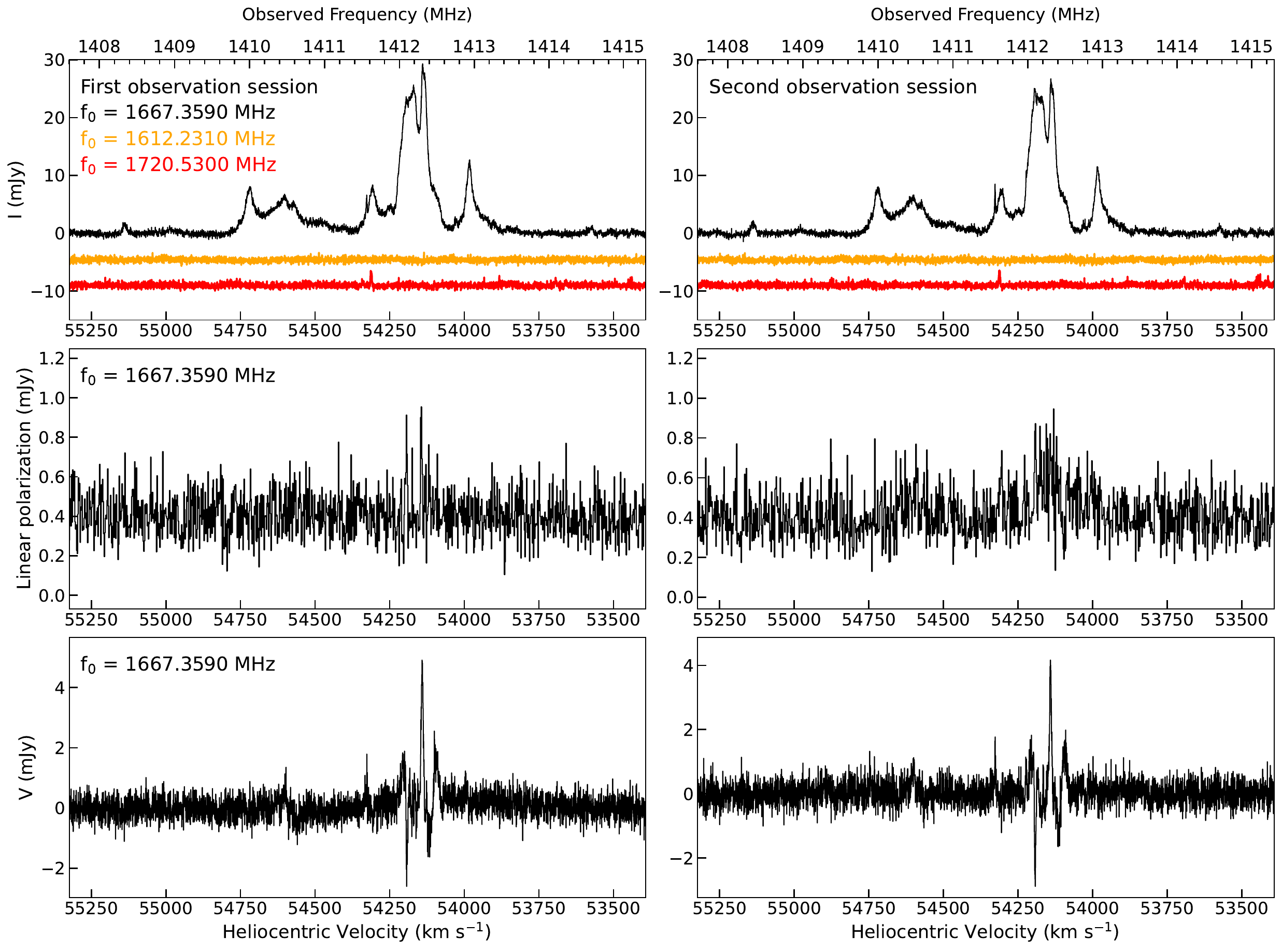}  
\caption{OH megamaser emission from {\it IRAS} 02524+2046 observed by FAST during two observation sessions. {\it Top panels}: Stokes $I$ spectra aligned by heliocentric velocity for the OH ground-state transitions. The vertical offsets are applied for clarity. {\it Middle panels}: Linear polarization spectra for the 1665~MHz and 1667~MHz OH megamaser lines. {\it Bottom panels}: Circular polarization spectra for both OH transitions. 
}
\label{2specobs}
\end{figure*}

Following standard calibration, we measured a linear polarization fraction $P_{src} = 9.6\% \pm 0.3\%$ at 1400~MHz for 3C\,286, consistent with the value $9.90\% \pm 0.003\%$ given by \citet{tl24} and the results of $\sim$9.3\% to 9.8\% reported by \citet{chl+24}, and we obtained the polarization angle $\chi = 36.4^\circ\pm0.8^\circ$. 
We calculated the Faraday rotation due to Earth's ionosphere with the package ionFR \citep[][]{ionfr}, and we then corrected the polarization angle to $27.2^\circ\pm0.9^\circ$ at 1400~MHz, consistent with the intrinsic polarization angle range of about $27.8^\circ-28.6^\circ$ at 1400~MHz given in Fig.~2 of \citet[][]{tl24}. This generally agrees with the $\chi$ values of $\sim 28^\circ$ to 32$^\circ$ at 1420\,MHz given by \citet{chl+24}.
Across our observed frequency range, the calibrated $V/I$ values are around zero, with absolute median and mean values of $0.03\%\pm0.01\%$ and $0.06\%\pm0.01\%$, respectively. We understand that 
the circular polarization of 3C\,286 should be zero at $L$ band \citep[e.g.,][]{tl24}, and any residuals should reflect the uncorrected leakage from $I$ to $V$ (i.e., $\alpha_\nu$$I_\nu$) in the 
calibration procedure above. This interpretation is supported by our subsequent analysis of Galactic $\hi$ 21~cm line observations.
We then further repaired the leakage terms $\alpha_\nu$$I_\nu$ by adding them to the calibrated $V$ spectrum, and we achieved a final circular polarization accuracy of about 0.01\%$-$0.08\% across $1050-1450$~MHz.

\begin{figure}[t]
\centering
\includegraphics[width=0.48\textwidth]{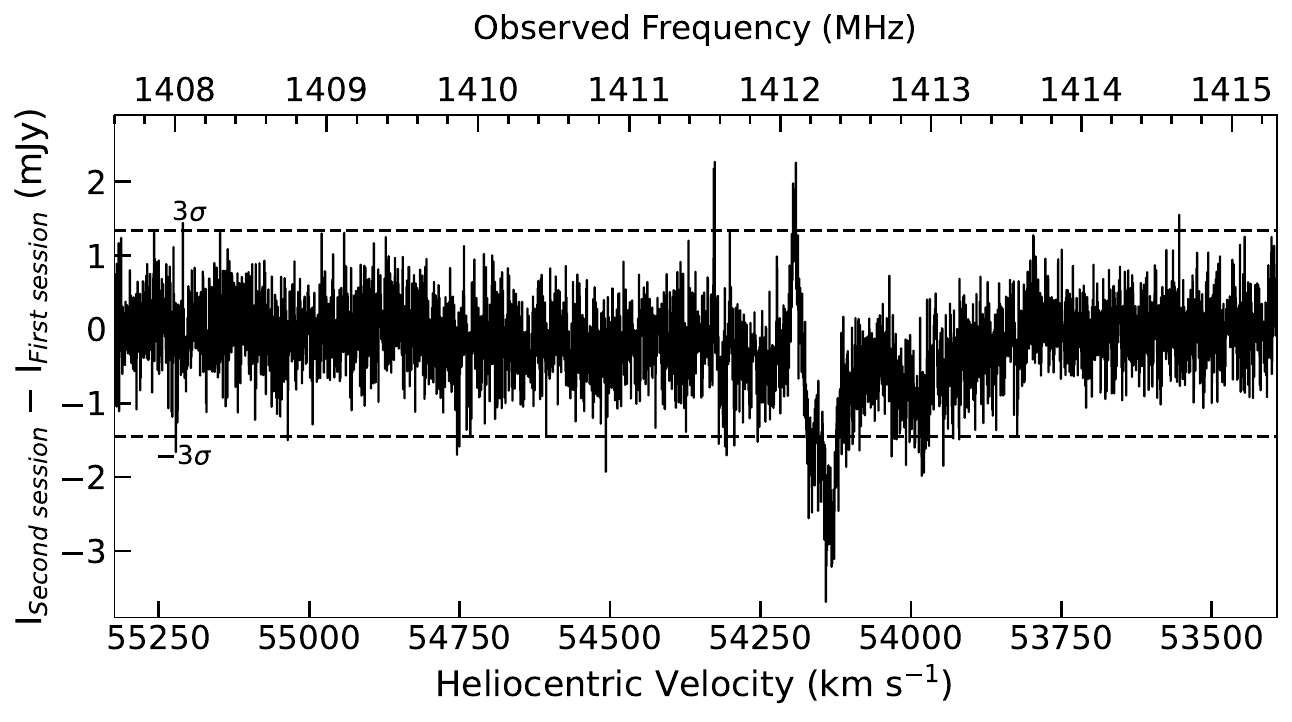}  
\caption{Difference in the Stokes $I$ spectrum between two observation sessions, indicating the one-day variation in the spectral lines.}
\label{OHvari}
\end{figure}

\begin{figure*}[t]
\centering
\includegraphics[width=0.68\textwidth]{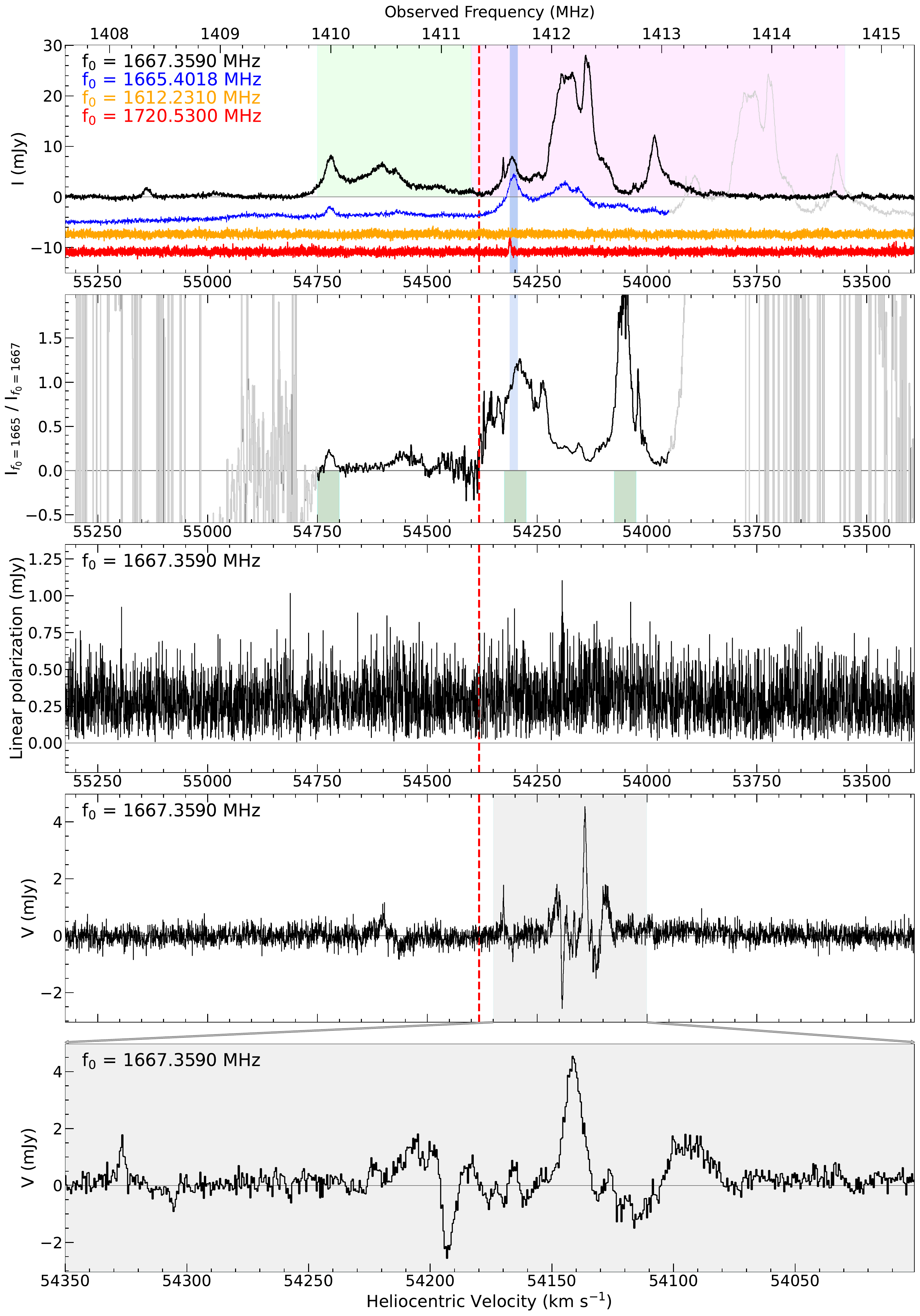}   
\caption{OH megamaser emissions from {\it IRAS} 02524+2046 observed by FAST for 110 minutes. {\it Top panel}: Stokes $I$ spectra of the OH ground state aligned by heliocentric velocity, with given vertical offsets for clarity. The dashed vertical line marks the heliocentric velocity $v_{\rm helio}= cz = 54\,382~{\rm km~s}^{-1}$ of {\it IRAS}~02524+2046 for a redshift $z=0.1814$ from optical spectrophotometry \citep{dg06}. The thick light blue line highlights the pair of 1667 and 1665~MHz OH megamaser lines, showing an unusual flux ratio reported by \citet{dg02} and \citet{mhe13b}. The fuchsia shaded area indicates the probable velocity range of the 1667 MHz OH megamaser emission lines, while the lime shaded area represents the velocity range in which the 1665 and 1667 MHz lines are likely mixed (see Sect.~\ref{sect:newresults}). {\it Upper middle panel}: Intensity ratio of the 1665~MHz to 1667~MHz lines. The dark green shaded areas indicate that the intensity ratio of the 1665~MHz to 1667~MHz transitions is potentially overestimated near $v_{\rm helio}\sim54\,050$~km~s$^{-1}$ and $v_{\rm helio}\sim54\,300$~km~s$^{-1}$, and underestimated near $v_{\rm helio}\sim54\,725$~km~s$^{-1}$.
{\it Middle panel}: Linear polarization spectrum at the heliocentric velocity for 1667.3590~MHz line. {\it Lower middle panel}: Stokes $V$ spectrum at the heliocentric velocity for 1667.3590~MHz line, showing some features from the 1665.4018~MHz transition. 
{\it Bottom panel}: Zoomed-in view of the Stokes $V$ spectral features.
}
\label{specobs}
\end{figure*}

To verify our calibration results, we analyzed Galactic $\hi$~21~cm line observations toward {\it IRAS}\,02524+2046 $(l=158.0^\circ, b=-33.3^\circ$) obtained with FAST during a 110-minute on-source integration.
After the baseline and standing waves were discounted \citep[see][and Sect.~\ref{swfitting} for examples]{jhh+23},
we derived the Stokes  $I$, $V$ and fractional polarization $V/I$ for the $\hi$~21~cm lines as shown in the lower panels of Fig.~\ref{galhi}. 
The fractional polarization $V/I$ ranges from about $-$0.06\% to 0.06\% across the line emission regions.
No similar shape of $V$ to the $I$ profiles implies that the uncompensated leakage of $I$ into $V$, if it exists, will be lower than 0.03\% near 1420.4~MHz. 
These results demonstrate FAST's capability to measure weak circular polarization with high precision. The consistent $V/I$ levels between calibrated 3C\,286 observations and the calibrated Galactic $\hi$~21~cm line observations suggest that the Mueller matrix values for FAST's central beam probably do not change significantly over timescales of weeks.

For timescales ranging from months to years, the Mueller matrix of FAST's 19-beam receiver central beam varies in time, as reported by \citet{chl+24}. When applying average Mueller matrix parameters to FAST observations from $2020-2022$, fractional circular polarization measurements exceeding 1.5\% can be considered reliable detections \citep{chl+24}. This performance is comparable to the reference-signal calibration method described in Sect.~\ref{sect:noise}.
These findings, together with our results, highlight the importance of regular calibrator observations for maintaining polarization calibration accuracy.

\section{The features of OH megamasers in {\it IRAS}\,02524+2046 observed by FAST}
\label{sect:results}

We applied the Mueller matrix calibration to the {\it IRAS}\,02524+2046 spectral data and obtained similar results from two independent FAST observations, as shown in Fig.~\ref{2specobs}. The one-day variation is apparent in some components (Fig.~\ref{OHvari}). After combining the two datasets, we derived the Stokes $I$ and $V$ spectra of the OH megamasers, as shown in Fig.~\ref{specobs}. Because the total on-source integration time is about 110 minutes by the super-sensitive FAST, we obtained the fine spectrum in Fig.~\ref{specobs} with a root mean square (RMS) noise of only about 0.3~mJy per channel at 0.48 km~s$^{-1}$ velocity resolution. Key findings include the detection of detailed OH megamaser features from {\it IRAS}\,02524+2046 and a high circular polarization reaching up to $\sim$16\% at certain frequencies. As we inspected the raw data in detail, these OH megamaser features detected by FAST are not caused by radio frequency interference.

\subsection{The new features of OH megamaser lines}
\label{sect:newresults}

The top panel of Fig.~\ref{specobs} shows the observed Stokes $I$ spectrum against the heliocentric velocity for the two main line transitions of the OH ground state at rest frequencies of 1665.4018~MHz and 1667.3590~MHz and the two satellite lines at the rest frequencies of 1612.2310~MHz and 1720.5300~MHz. 
The 1612 MHz spectrum shows no detectable features, and the RMS level of $\sim$0.3~mJy demonstrates FAST's exceptional sensitivity. 
A very narrow emission line feature in the 1720~MHz spectrum was detected with a peak flux density of $2.4\pm0.1$~mJy, a line center $54\,312.0\pm0.1$~km~s$^{-1}$ in the heliocentric frame, and a full width at half maximum (FWHM) line width $4.6\pm0.3$~km~s$^{-1}$. This feature appears in the two independent FAST observation sessions (see Fig.~\ref{2specobs}). 
The line width is smaller than the typical 10~km~s$^{-1}$ width of OH megamaser components \citep[][]{le08}. 
To date, OH satellite lines in the $L$ band have only been detected in seven galaxies: Arp\,220 \citep{bh87,mhe13b}, III ZW\,35 \citep{bhh89,mhe13b}, {\it IRAS}\,17207$-$0014 \citep{bhh89,mhe13b}, Arp\,299 (IC\,694) and Mrk\,231 \citep{bhh92}, {\it IRAS}\,10173+0829, and {\it IRAS}\,15107+0724 \citep{mhe13b}.

The dominant 1665~MHz and 1667~MHz OH megamaser features appear within $v_{\rm helio} \sim 53\,800 - 54\,850~{\rm km~s}^{-1}$, consistent with previous single-dish \citep[][]{dg02,mhe13b,wsz+23} and interferometry \citep[][]{pwz+20} observations. 
However, significant flux density discrepancies exist among different studies. When we take the strongest feature near $v_{\rm helio} \sim 54\,150$~km~s$^{-1}$ as an example, the flux density values given by different works are $\sim$28~mJy (this work), $\sim$30~mJy~\citep[][]{wsz+23}, $\sim$80~mJy~\citep[][]{mhe13b}, and $\sim$40~mJy~\citep[][]{dg02}. 
The OH megamaser variability likely contributes to these differences.
This variability was first discovered by \citet{dg02v} and is commonly interpreted as a result of interstellar scintillation \citep[e.g.,][]{dg02v, wsz+23}, although not all of the components present apparent variation. Intrinsic changes in the physical conditions of the maser environment \citep[][]{mrh15, hag16} might also account for the variability.
For {\it IRAS}\,02524+2046, the strong variability of OH megamasers across multiple spectral components has been documented by Arecibo observations. Notably, the RMS spectrum reveals significant day-to-day variations~\citep[][]{darl05, darl07}. 
The short-term variability is also evident in our data. As shown in Fig.~\ref{OHvari}, day-to-day changes are noticeable in some emission components centered from $v_{\rm helio} \sim 54\,000$~km~s$^{-1}$ to 54\,250~km~s$^{-1}$.

In comparison to previous works, the FAST spectrum presents more detailed emission line features resting on its high sensitivity and high spectral resolution. For instance, a prominent narrow emission line component from the 1667~MHz transition appears near $v_{\rm helio} \sim 54\,327$~km~s$^{-1}$, which has 
a peak flux density of $3.9\pm0.2$~mJy, a line center of $54\,326.96\pm0.06$~km~s$^{-1}$, and a FWHM line width of $2.6\pm0.2$~km~s$^{-1}$.
The line width is narrower than that of typical OH megamaser components \citep[10~km~s$^{-1}$,][]{le08} and is broader than Galactic star-forming region OH masers \citep[$\lesssim$~1~km~s$^{-1}$, e.g.,][]{cgp13,cgp14}.
This narrow component is unlikely to originate from RFI, as it was consistently detected across multiple independent observational datasets: in both sessions of our FAST observations, and in the separate observations conducted by \citet{dg02} and \citet{mh13} with Arecibo.
Outside this velocity range, there are two emission line features.
One feature is a new $\sim$3$\sigma$ detection at  $v_{helio}\sim53\,580$~km~s$^{-1}$, most likely from 1667~MHz OH lines considering its observed frequency.
The other is a $>$5$\sigma$ feature at $v_{helio}\sim55\,140$~km~s$^{-1}$, which was previously reported by \citet[][see the upper right panel of their Fig.~1]{dg02}, and which corresponds to a heliocentric velocity of $\sim$54\,725~km~s$^{-1}$ for 1665~MHz OH transition.

The detection of the 1665~MHz line near the heliocentric velocity of $\sim$54\,725~km~s$^{-1}$ provides crucial insight into the unusual flux ratio of the pair of 1667 MHz and 1665 MHz~lines previously reported~\citep{dg02,mhe13b}. 
In the Arecibo survey results for 50 OH megamaser galaxies, all the identified 1665~MHz OH megamaser lines showed much stronger 1667~MHz counterparts \citep[][]{dg01a,dg01b,dg02}. 
As shown by the blue spectrum in Fig.~\ref{specobs}, the weak feature far left near $v_{\rm helio}\sim55\,140$~km~s$^{-1}$ in the black spectrum should be attributed to the 1665~MHz transition at the heliocentric velocity $\sim54\,725$~km~s$^{-1}$, where a much stronger emission line component appears in the black spectrum. 
The emission line component near $v_{\rm helio}\sim54\,725$~km~s$^{-1}$ in the black spectrum should not be simply attributed to the 1665~MHz OH megamasers alone, as done by e.g., \citet{dg02}, \citet{mhe13b}, and \citet{pwz+20}, but is likely to be blended emission from the 1667~MHz and 1665~MHz transitions.
We adopted a median $R_H$ value of 5.9 and a typical interquartile range of 3.2$-$8.4 from the OH megamaser sample of \citet[][]{dg02b}. Using this typical $R_H$ value and the integrated intensity of the 1665~MHz megamasers at $v_{helio}\sim55\,140$~km~s$^{-1}$, we estimated the contribution of the 1667~MHz line to the emission at $v_{\rm helio}\sim54\,725$~km~s$^{-1}$. This allowed us to subsequently estimate an $R_H$ value of 2.4$^{+2.8}_{-0.9}$ for the emissions near $v_{\rm helio}\sim54\,300$~km~s$^{-1}$.
Consequently, the previously reported anomalous line ratio near $v_{\rm helio}\sim54\,300$~km~s$^{-1}$ can be naturally explained by the blending of both lines at the same velocities. 
Therefore, the intensity ratio of the 1665~MHz to 1667~MHz transitions, shown in the upper middle panel of Fig.~\ref{specobs}, is likely overestimated near $v_{\rm helio}\sim54\,300$~km~s$^{-1}$ and underestimated near $v_{\rm helio}\sim54\,725$~km~s$^{-1}$. A potential overestimation may also exist near $v_{\rm helio}\sim54\,050$~km~s$^{-1}$, although the individual maser lines are difficult to distinguish in this region.
Our results suggest that the 1667~MHz OH megamaser emission
from {\it IRAS}\,02524+2046 extends beyond $v_{\rm helio} \sim$~54\,400~km~s$^{-1}$, with significant components at higher velocities, as shown in Fig.~\ref{specobs}.
Additionally, we note that the new interpretation reveals that the 1667~MHz OH megamaser emissions are distributed on either side of the systemic velocity of {\it IRAS}\,02524+2046 \citep[$v_{\rm helio}=54\,382~{\rm km~s}^{-1}$, determined from optical spectrophotometry,][]{dg06}. This differs from previous results, which suggested emissions on only one side \citep[e.g.,][]{dg02,mhe13b}.

The linear polarization spectrum of the 1665~MHz and 1667~MHz OH megamasers is shown in the middle panel of Fig.~\ref{specobs}. We detect 
a polarization feature with $\sim3.6\sigma$ significance at $v_{\rm helio} \sim$~54\,190~km~s$^{-1}$. This feature has a polarized intensity of $\sim$0.54~mJy compared to the total intensity peak of $\sim$24 mJy at the same velocity, corresponding to a linear polarization degree of $\sim$2.3\%. 
The origin of the linear polarization observed in the OH megamasers remains unclear. An instrumental leakage of Stokes $I$ to $Q$ and/or $U$ is unlikely, as the intense OH emission features around $v_{\rm helio} \sim$~54\,150~km~s$^{-1}$ in the $I$ spectrum would otherwise also imprint a detectable signature on the linear polarization spectrum. Instead, we speculate that the linear polarization might originate from a $\pi$ component, a phenomenon detected in approximately 16\% of Galactic OH masers, including ground-state and excited-state transitions~\citep[][]{gcm+15}. 
No other polarization features with a significance above 3$\sigma$ are present in the linear polarization spectrum.

The Stokes $V$ spectrum for the main OH ground-state transitions is shown in the lower two panels of Fig.~\ref{specobs}. In addition to the three prominent features discussed by \citet{mhe13b}, our FAST observations reveal new spectral components that appear as peaks and dips within the velocity range $v_{\rm helio}\sim$~54\,070 $-$ 54\,350~km~s$^{-1}$. 
While Zeeman splitting remains the primary mechanism for Stokes $V$ features in OH masers, non-Zeeman effects might contribute as well. Theoretical studies indicated that the linear-to-elliptical polarization conversion and Faraday rotation in maser-emitting clouds can produce antisymmetric $V$ profiles, which particularly affect weakly split interstellar masers of SiO, H$_2$O, and CH$_3$OH~\citep{wat09}. For OH masers in star-forming regions, observed $V$ profiles typically reflect Zeeman splitting, but might be modified by magnetic field gradients, velocity gradients, and radiative transfer effects~\citep[e.g.,][]{nw90}. The situation becomes more complex for OH megamasers, where line couplings between Doppler-shifted transitions (e.g., the 1665~MHz and 1667~MHz lines) in the velocity gradient conditions of gas clouds might additionally affect the $V$ profiles, although the observed $V$ profiles of OH megamasers are also thought to be Zeeman dominated~\citep[][]{rqh08,mh13}. Following established methods~\citep[][]{rqh08,mh13}, we analyzed the observed Stokes $I$ and $V$ spectra to extract magnetic field information.

\begin{figure*}
 \centering
  \includegraphics[width=0.9\textwidth]{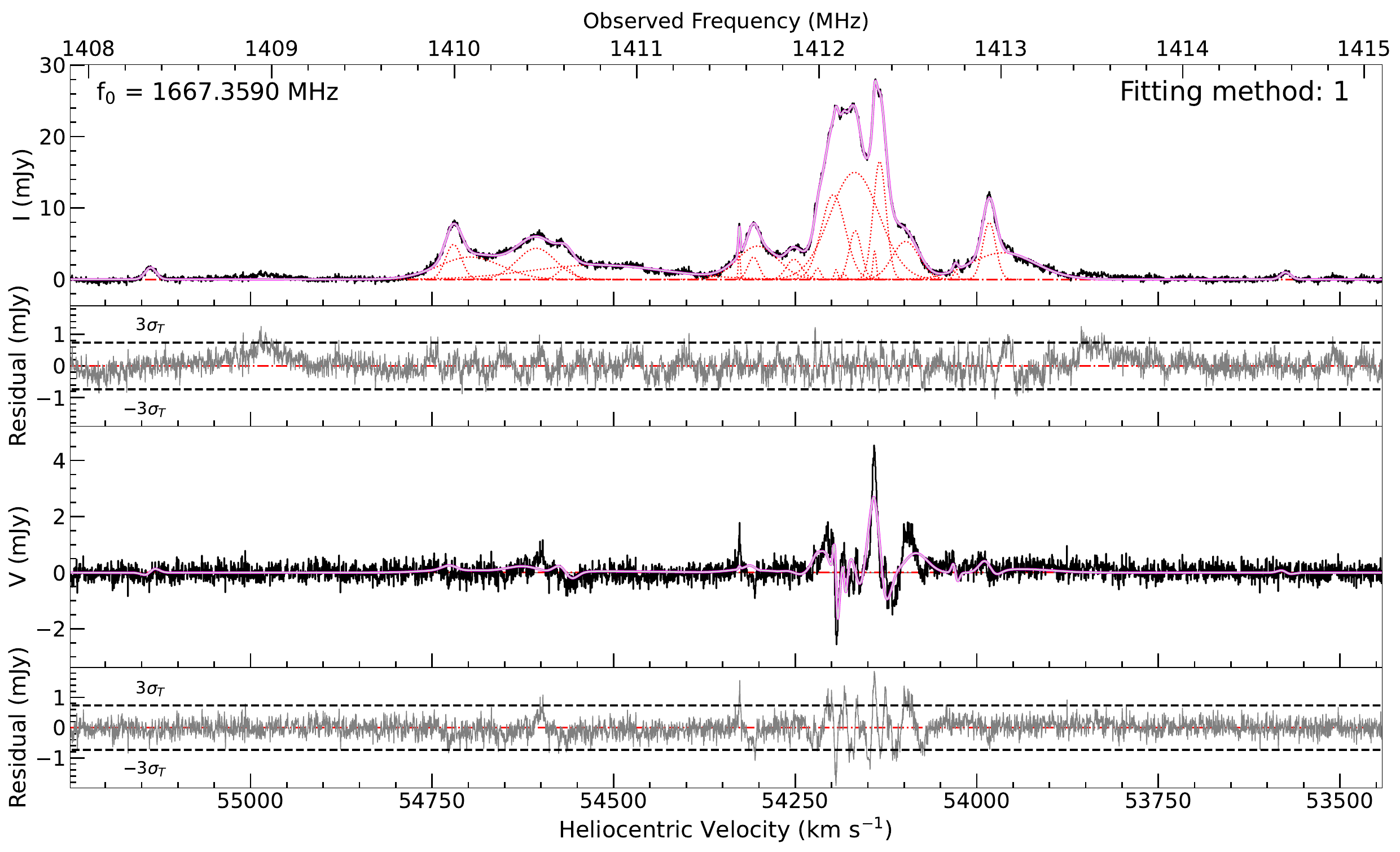}\\
 \vspace{0.25cm}
 \includegraphics[width=0.9\textwidth]{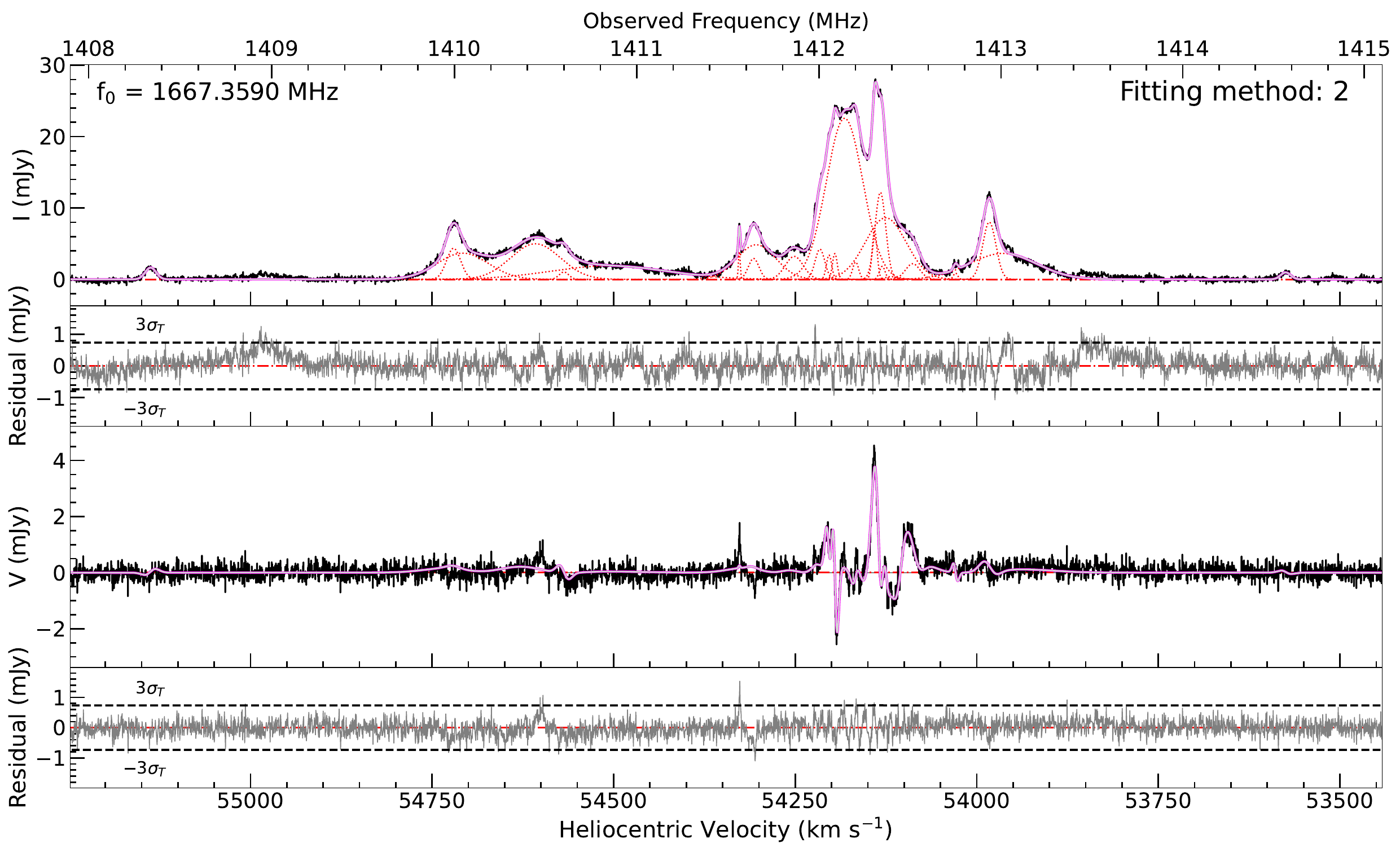}
  \caption{Zeeman-splitting analysis of OH megamaser lines from {\it IRAS} 02524+2046. {\it Upper panels}: Stokes $I$ and $V$ spectral fits obtained using the method described in Sect~\ref{sect:fitting1} (fitting method 1), while {\it the lower panels} display results from another approach in Sect.~\ref{sect:fitting2} (fitting method 2). For heliocentric velocity calculations, we adopted a rest frequency of $\nu_0$~=~1667.3590~MHz, although some spectral features originate from the 1665.4018~MHz OH transition. }
 \label{gaufit}
\end{figure*}

\begin{table*}
\caption{Zeeman-splitting analysis results obtained from fitting the Stokes $I$ and $V$ spectra of OH megamasers in {\it IRAS}\,02524+2046 with the method described in Sect~\ref{sect:fitting1}.} 
\label{tabfit1}    
\tabcolsep 3pt 
\centering                         
\begin{tabular}{lccccccc}       
\hline\hline                
Gaussian & $S$ & $\nu$   &  $\Delta \nu$ &  $B_{//}$ &  $B_{//}^{1st}$ &  $B_{//}^{2nd}$  &   Heliocentric velocity\\
         & (mJy) & (MHz)    &  (MHz) &  (mG)          &    (mG)   &   (mG)  &  (km~s$^{-1}$)  \\
   (1)   &  (2) & (3)   & (4)   & (5)             &     (6)   &   (7)     & (8) \\
\hline  
1  &  1.65 $\pm$ 0.08& 1408.3385 $\pm$ 0.0016 & 0.0299 $\pm$ 0.0016&                  &                 &                   &  54720.96 $\pm$ 0.40$^\dagger$ \\
2  &  4.88 $\pm$ 0.12& 1409.9954 $\pm$ 0.0007 & 0.0436 $\pm$ 0.0011&                  &                 &                   &  54304.36 $\pm$ 0.18$^\dagger$ \\
3  &  3.14 $\pm$ 0.16& 1410.0873 $\pm$ 0.0101 & 0.1700 $\pm$ 0.0077&                  &                 &                   &  54281.29 $\pm$ 2.55$^\dagger$ \\
4  &  4.35 $\pm$ 0.18& 1410.4488 $\pm$ 0.0032 & 0.1058 $\pm$ 0.0055&    3.6 $\pm$ 0.9$^*$ &   1.8 $\pm$ 1.0$^*$ &    5.5 $\pm$ 1.1$^*$  &  54190.54 $\pm$ 0.81$^\dagger$ \\  
5  &  1.65 $\pm$ 0.16& 1410.6081 $\pm$ 0.0024 & 0.0400 $\pm$ 0.0035&    6.5 $\pm$ 1.3$^*$ &   5.8 $\pm$ 1.7$^*$ &    7.5 $\pm$ 1.6$^*$  &  54150.56 $\pm$ 0.60$^\dagger$ \\ 
6  &  2.02 $\pm$ 0.04& 1410.7831 $\pm$ 0.0299 & 0.3664 $\pm$ 0.0243&                  &                 &                   &  54106.66 $\pm$ 7.51$^\dagger$ \\
7  &  3.97 $\pm$ 0.19& 1411.5623 $\pm$ 0.0003 & 0.0054 $\pm$ 0.0003&                  &                 &                   &  54326.98 $\pm$ 0.07  \\
8  &  3.12 $\pm$ 0.20& 1411.6399 $\pm$ 0.0010 & 0.0299 $\pm$ 0.0016&                  &                 &                   &  54307.52 $\pm$ 0.24  \\
9  &  4.65 $\pm$ 0.18& 1411.6620 $\pm$ 0.0035 & 0.1107 $\pm$ 0.0047&                  &                 &                   &  54301.97 $\pm$ 0.89  \\
10 &  2.81 $\pm$ 0.44& 1411.8626 $\pm$ 0.0037 & 0.0424 $\pm$ 0.0039&                  &                 &                   &  54251.67 $\pm$ 0.92  \\
11 &  1.55 $\pm$ 0.30& 1411.9924 $\pm$ 0.0018 & 0.0168 $\pm$ 0.0027&                  &                 &                   &  54219.12 $\pm$ 0.45  \\
12 & 11.84 $\pm$ 3.51& 1412.0781 $\pm$ 0.0063 & 0.0659 $\pm$ 0.0099&   11.4 $\pm$ 0.5 &  11.7 $\pm$ 0.6 &   10.3 $\pm$ 0.6  &  54197.62 $\pm$ 1.59  \\  
13 &  1.36 $\pm$ 0.29& 1412.0935 $\pm$ 0.0017 & 0.0094 $\pm$ 0.0023&   16.0 $\pm$ 1.3 &  13.0 $\pm$ 1.2 &   26.8 $\pm$ 2.5  &  54193.76 $\pm$ 0.44  \\   
14 &  0.79 $\pm$ 0.39& 1412.1386 $\pm$ 0.0033 & 0.0105 $\pm$ 0.0049&   17.6 $\pm$ 2.4 &   7.4 $\pm$ 8.0 &   14.7 $\pm$ 2.8  &  54182.46 $\pm$ 0.84  \\  
15 & 14.99 $\pm$ 2.56& 1412.1953 $\pm$ 0.0378 & 0.1349 $\pm$ 0.0181&$-$11.5 $\pm$ 0.5 &$-$11.1 $\pm$ 0.8&$-$11.5 $\pm$ 0.6  &  54168.24 $\pm$ 9.47  \\   
16 &  6.81 $\pm$ 1.40& 1412.1989 $\pm$ 0.0026 & 0.0335 $\pm$ 0.0032&    9.2 $\pm$ 0.6 &  10.4 $\pm$ 0.9 &    8.0 $\pm$ 0.7  &  54167.35 $\pm$ 0.64  \\  
17 &  3.95 $\pm$ 0.29& 1412.3056 $\pm$ 0.0004 & 0.0091 $\pm$ 0.0007&                  &                 &                   &  54140.62 $\pm$ 0.10  \\
18 & 16.52 $\pm$ 1.51& 1412.3329 $\pm$ 0.0013 & 0.0354 $\pm$ 0.0015&    6.8 $\pm$ 0.2 &   6.3 $\pm$ 0.3 &    7.0 $\pm$ 0.3  &  54133.77 $\pm$ 0.32  \\  
19 &  5.30 $\pm$ 1.59& 1412.4760 $\pm$ 0.0155 & 0.0733 $\pm$ 0.0078&$-$13.4 $\pm$ 1.1 &$-$11.5 $\pm$ 1.3&$-$14.6 $\pm$ 1.3  &  54097.91 $\pm$ 3.89  \\  
20 &  1.01 $\pm$ 0.12& 1412.7497 $\pm$ 0.0017 & 0.0130 $\pm$ 0.0019&    7.7 $\pm$ 2.0 &   9.5 $\pm$ 2.4 &    5.6 $\pm$ 2.4  &  54029.34 $\pm$ 0.43  \\  
21 &  7.97 $\pm$ 0.08& 1412.9354 $\pm$ 0.0004 & 0.0358 $\pm$ 0.0005&    2.1 $\pm$ 0.4 &   1.9 $\pm$ 0.5 &    2.2 $\pm$ 0.5  &  53982.86 $\pm$ 0.10  \\  
22 &  3.73 $\pm$ 0.05& 1413.0107 $\pm$ 0.0024 & 0.1782 $\pm$ 0.0022&                  &                 &                   &  53964.01 $\pm$ 0.61  \\
23 &  0.98 $\pm$ 0.08& 1414.5672 $\pm$ 0.0026 & 0.0277 $\pm$ 0.0026&                  &                 &                   &  53574.74 $\pm$ 0.64  \\
\hline                                   
\end{tabular}
\tablefoot{Column (1) numbers the Gaussian components. Columns (2)$-$(4) present the fit parameters (peak flux density, line center, and line width) with their 1$\sigma$ uncertainties. Column (5) shows the derived magnetic field strength and its uncertainty, where values annotated with an asterisk designate components attributed to the 1665~MHz OH transition (using a splitting coefficient of 3.270~Hz~$\mu$G$^{-1}$). When these emissions are instead assumed to be dominated by 1667~MHz OH transition using a splitting coefficient of 1.964~Hz~$\mu$G$^{-1}$, the derived values are 6.1~$\pm$~1.5~mG and 10.9~$\pm$~2.2~mG for Gaussian components 4 and 5, respectively. Columns (6) and (7) separately list magnetic field values obtained from fitting the first- and second-epoch observations independently. Column (8) gives the heliocentric velocities, and the dagger denotes values corresponding to the assumed 1665~MHz OH maser components.}
\end{table*}

\subsection{The Zeeman splittings of OH megamaser lines}
\label{sect:fitting}

Following \citet{rqh08} and \citet{mh13}, we assumed that the Zeeman splitting of every OH megamaser component is smaller than the line width. The $V$ spectrum can then be expressed as
\begin{align}
  \label{eq1}
V = \sum\limits_{i=1}^n (\frac{\nu}{\nu_{0i}})(\frac{dI_i}{d\nu})\frac{b_i}{2}B_{// i}+C_{I2V},
\end{align}
where $\nu$ is the observing frequency, $\nu_{0i}$ is the rest
frequency of the OH transition for the $i$th OH megamaser line component,
$\nu/\nu_{0i}$ accounts for the frequency compression of the redshifted
line, $I_i$ is the total intensity profile of the $i$th OH megamaser
component, $b_i$ is the splitting coefficient, which is 
1.964~Hz~$\mu$G$^{-1}$ for the 1667.3590~MHz line and
3.270~Hz~$\mu$G$^{-1}$ for the 1665.4018~MHz line \citep[][]{hei93,robi08}, and $B_{//i}$ is the strength of the line-of-sight magnetic field with a positive value for magnetic fields pointing away from the observer. The parameter $C_{I2V}$ was originally used to quantify the uncompensated leakage from Stokes $I$ to $V$.  However, in our analysis, the fitted $C_{I2V}$ represents a composite signal, which includes both the genuine uncompensated $I$-to-$V$ leakage and residual standing wave contamination. This is because residual components persist despite the modeling and subtraction of the broad-scale slope of the standing wave, as illustrated in Fig.~\ref{swfitting}.

As discussed in Sect.~\ref{sect:newresults}, all the OH emission lines with
$v_{\rm helio}\lesssim54\,400$~km~s$^{-1}$ stem from the 1667~MHz OH megamasers, and we hence adopted $b_i=$~1.964~Hz~$\mu$G$^{-1}$ in the fitting.
As discussed above, the 1667 and 1665~MHz OH emission lines are likely mixed for the emission feature near $v_{\rm helio} \sim 54\,725$~km~s$^{-1}$. For the velocity range $v_{\rm helio} \sim$~54\,400~$-$~54\,700~km~s$^{-1}$, the 1667 and 1665~MHz lines might also be mixed, although observational support remains insufficient for two reasons: (1) as shown in the upper panel of Fig.~\ref{specobs}, no obvious 1665 MHz emissions are detected in the range $v_{\rm helio} \sim 54\,800~$–$~55\,100$ km s$^{-1}$, and (2) the intensity ratio of the 1665 MHz to 1667 MHz lines in $v_{\rm helio} \sim 54\,100~$–$~54\,200$ km s$^{-1}$ is not anomalous. At least two possibilities exist: (1) the OH emissions in $v_{\rm helio} \sim 54\,400~$–$~54\,700$ km s$^{-1}$ are dominated by the 1665 MHz transition, or (2) these emissions are primarily from the 1667 MHz transition, but lack counterparts of 1665 MHz emissions at similar velocities. Based on the available FAST single-dish data, it remains challenging to distinguish between these scenarios. If the emissions in the velocity range $v_{\rm helio} \sim$~54\,400~$-$~54\,725~km~s$^{-1}$ result from a combination of the 1665~MHz and 1667~MHz lines, the magnetic field fitting will involve derivatives of the Stokes $I$ emission with uncertain contributions from each line, leading to a complex error propagation.
For simplicity, we first adopted $b_i = 3.270$ Hz $\mu$G$^{-1}$ for OH emission lines with $v_{\rm helio} > 54,400$ km s$^{-1}$, and we then used $b_i = 1.964$ Hz $\mu$G$^{-1}$ to repeat the fitting for further discussion.

In the following, we apply two methods to fit the observed $I$ and $V$ spectra. The first method follows the procedures described by \citet{rqh08} and \citet{mh13} and first decomposes the $I$ spectrum into multiple Gaussians and then examines the possible Zeeman splittings. The second method is to fit the $I$ and $V$ spectra simultaneously with multiple Gaussian components.

\begin{table*}
\caption{Same as Table~\ref{tabfit1}: Analysis results obtained using the method described in Sect~\ref{sect:fitting2}.} 
\label{tabfit2}    
\tabcolsep 3pt  
\centering                         
\begin{tabular}{lccccccc}       
\hline\hline                
Gaussian & $S$ & $\nu$   &  $\Delta \nu$ &  $B_{//}$ &  $B_{//}^{1st}$ &  $B_{//}^{2nd}$  &   Heliocentric velocity\\
         & (mJy) & (MHz)    &  (MHz) &  (mG)          &    (mG)   &   (mG)  &  (km~s$^{-1}$)  \\
   (1)   &  (2) & (3)   & (4)   & (5)             &     (6)   &   (7)     & (8) \\
\hline  
 1 &  1.65 $\pm$ 0.07 & 1408.3387 $\pm$ 0.0014 & 0.0298 $\pm$ 0.0014&                      &                      &           &  54720.91 $\pm$ 0.35$^\dagger$ \\
 2 &  4.35 $\pm$ 0.11 & 1409.9958 $\pm$ 0.0007 & 0.0388 $\pm$ 0.0010&                      &                      &           &  54304.28 $\pm$ 0.17$^\dagger$ \\
 3 &  3.68 $\pm$ 0.10 & 1410.0460 $\pm$ 0.0046 & 0.1406 $\pm$ 0.0042&                      &                      &           &  54291.67 $\pm$ 1.16$^\dagger$ \\  
 4 &  4.99 $\pm$ 0.20 & 1410.4444 $\pm$ 0.0027 & 0.1361 $\pm$ 0.0046&    4.0 $\pm$ 0.8$^*$ &    2.8 $\pm$ 1.0$^*$ &    5.7 $\pm$ 1.1$^*$&  54191.66 $\pm$ 0.67$^\dagger$ \\  
 5 &  1.04 $\pm$ 0.09 & 1410.6005 $\pm$ 0.0020 & 0.0264 $\pm$ 0.0025&    7.3 $\pm$ 1.7$^*$ &    7.8 $\pm$ 2.4$^*$ &    7.0 $\pm$ 1.9$^*$&  54152.48 $\pm$ 0.50$^\dagger$ \\  
 6 &  1.90 $\pm$ 0.04 & 1410.8472 $\pm$ 0.0280 & 0.3175 $\pm$ 0.0217&                      &                      &           &  54090.58 $\pm$ 7.03$^\dagger$ \\
 7 &  3.94 $\pm$ 0.17 & 1411.5624 $\pm$ 0.0002 & 0.0051 $\pm$ 0.0003&                      &                      &           &  54326.96 $\pm$ 0.06 \\
 8 &  2.93 $\pm$ 0.17 & 1411.6410 $\pm$ 0.0009 & 0.0290 $\pm$ 0.0015&                      &                      &           &  54307.25 $\pm$ 0.24 \\
 9 &  4.85 $\pm$ 0.16 & 1411.6553 $\pm$ 0.0028 & 0.1068 $\pm$ 0.0038&                      &                      &           &  54303.65 $\pm$ 0.69 \\
10 &  3.22 $\pm$ 0.16 & 1411.8640 $\pm$ 0.0026 & 0.0495 $\pm$ 0.0027&                      &                      &           &  54251.30 $\pm$ 0.65 \\
11 &  4.23 $\pm$ 0.39 & 1412.0049 $\pm$ 0.0010 & 0.0266 $\pm$ 0.0017&                      &                      &           &  54215.97 $\pm$ 0.25 \\
12 &  3.40 $\pm$ 0.21 & 1412.0550 $\pm$ 0.0009 & 0.0144 $\pm$ 0.0009&   11.7 $\pm$ 1.0 &   14.6 $\pm$ 1.5 &   11.0 $\pm$ 1.2&  54203.43 $\pm$ 0.23 \\   
13 &  3.62 $\pm$ 0.26 & 1412.0870 $\pm$ 0.0006 & 0.0130 $\pm$ 0.0005&   17.1 $\pm$ 1.2 &   14.9 $\pm$ 1.2 &   20.0 $\pm$ 2.0&  54195.39 $\pm$ 0.15 \\   
14 & 22.56 $\pm$ 0.29 & 1412.1406 $\pm$ 0.0027 & 0.1008 $\pm$ 0.0026&                      &                      &           &  54181.96 $\pm$ 0.69 \\
15 &  2.43 $\pm$ 0.15 & 1412.2011 $\pm$ 0.0008 & 0.0165 $\pm$ 0.0010& $-$4.4 $\pm$ 0.9 & $-$6.1 $\pm$ 1.4 & $-$3.5 $\pm$ 1.1&  54166.80 $\pm$ 0.20 \\  
16 &  6.89 $\pm$ 0.54 & 1412.3052 $\pm$ 0.0006 & 0.0134 $\pm$ 0.0006&                      &                      &           &  54140.70 $\pm$ 0.15 \\
17 & 12.18 $\pm$ 0.84 & 1412.3370 $\pm$ 0.0010 & 0.0292 $\pm$ 0.0006&   18.4 $\pm$ 2.1 &   19.5 $\pm$ 4.6 &   17.0 $\pm$ 1.5&  54132.74 $\pm$ 0.25 \\  
18 &  1.83 $\pm$ 0.71 & 1412.3473 $\pm$ 0.0010 & 0.0155 $\pm$ 0.0013&                      &                      &           &  54130.16 $\pm$ 0.24 \\
19 &  8.63 $\pm$ 0.83 & 1412.3662 $\pm$ 0.0058 & 0.1116 $\pm$ 0.0024&$-$24.5 $\pm$ 3.8 &$-$21.1 $\pm$ 3.3 &$-$32.8 $\pm$ 9.1&  54125.42 $\pm$ 1.45 \\  
20 &  0.39 $\pm$ 0.51 & 1412.3964 $\pm$ 0.0020 & 0.0329 $\pm$ 0.0023&                      &                      &           &  54117.85 $\pm$ 0.49 \\
21 &  2.20 $\pm$ 0.21 & 1412.5148 $\pm$ 0.0031 & 0.0398 $\pm$ 0.0030&   20.6 $\pm$ 3.2 &   16.5 $\pm$ 3.0 &   29.8 $\pm$ 6.3&  54088.20 $\pm$ 0.77 \\  
22 &  0.87 $\pm$ 0.11 & 1412.7501 $\pm$ 0.0014 & 0.0119 $\pm$ 0.0014&    8.4 $\pm$ 2.4 &   11.9 $\pm$ 3.7 &    5.8 $\pm$ 2.5&  54029.24 $\pm$ 0.34 \\   
23 &  8.02 $\pm$ 0.07 & 1412.9356 $\pm$ 0.0003 & 0.0362 $\pm$ 0.0004&    2.0 $\pm$ 0.4 &    1.9 $\pm$ 0.5 &    2.2 $\pm$ 0.5&  53982.81 $\pm$ 0.09 \\  
24 &  3.64 $\pm$ 0.04 & 1413.0112 $\pm$ 0.0022 & 0.1838 $\pm$ 0.0021&                      &                      &           &  53963.86 $\pm$ 0.56 \\
25 &  0.98 $\pm$ 0.07 & 1414.5674 $\pm$ 0.0023 & 0.0277 $\pm$ 0.0023&                      &                      &           &  53574.70 $\pm$ 0.57 \\
\hline                                   
\end{tabular}
\tablefoot{Same as the notes of Table~\ref{tabfit1}. In Column (5), the values annotated with an asterisk designate components attributed to the 1665~MHz OH transition. When these emissions are instead assumed to be dominated by the 1667~MHz OH transition, the derived values are 6.6$\pm$1.3~mG and 12.1$\pm$2.8~mG.}
\end{table*}

\subsubsection{Fitting using the first method}
\label{sect:fitting1}

Following the method described by \citet[][]{robi08} and \citet{mh13}, we began by solving equation~(\ref{eq1}) by decomposing the Stokes $I$ spectrum into multiple Gaussian components. 
Using the 18 Gaussian components given by \citet{mh13} as initial guesses, we identified five additional Gaussian components required to fit the features revealed by FAST observations. These include two emission line components near $v_{\rm helio}\sim55\,140$~km~s$^{-1}$ and
$\sim53\,580$~km~s$^{-1}$, one narrow emission line near $v_{\rm helio}\sim54\,327$~km~s$^{-1}$, and two components for resolved peaks around
$v_{\rm helio}\sim54\,180$~km~s$^{-1}$ in the Stokes $I$ spectrum. 
The fitting result of Stokes $I$ is shown in the upper panel of Fig.~\ref{gaufit}. Then, these Gaussian components were used to derive $dI/d\nu$ to fit the $V$ spectrum \citep[e.g.,][]{rqh08,mh13}, as presented in Fig.~\ref{gaufit}. The corresponding parameters are listed in Table~\ref{tabfit1}.

Consistent with \citet{mh13}, we obtained satisfactory fits for the Stokes $I$ spectrum of OH megamasers in {\it IRAS}\,02524+2046, but encountered difficulties in fitting the complex $V$ profile features. Significant residuals ($>$ 3$\sigma_T$, see Fig.~\ref{gaufit}) persist in the velocity range $v_{helio}\sim$~54\,050 $-$ 54\,250~km~s$^{-1}$, despite attempts to optimize initial guesses for a simultaneous $I$ and $V$ spectrum fitting.
\citet{mh13} reported confident magnetic field detections (+12.27 to +23.88 mG) for five narrow components (Gaussians 3, 8, 9, 12, and 13 in their Table 7). We reproduced four of these components, but did not detect their Gaussian component 13 because the local peak at 1412.5320~MHz observed by \citet{mh13} is absent in our FAST spectra. 
Our magnetic field measurements agree with theirs within 3$\sigma$ uncertainties for components 3, 8, and 9 in their Table 7. However, for their component 12 \citep[$B_{//}=$~13.65$\pm$1.07~mG in][]{mh13}, we measure a smaller field (6.8$\pm$0.2~mG, i.e., our component 18 in Table~\ref{tabfit1}). The discrepancy in the fitting results for this component is likely attributable to the intrinsic variability of the magnetic field. Moreover, these findings support a predominantly positive orientation of the magnetic field within the OH megamasers of {\it IRAS}\,02524+2046.

\subsubsection{Fitting using the second method}
\label{sect:fitting2}

An optimal decomposition of multiple OH megamaser components should accurately reproduce the $I$ spectrum and detailed $V$ spectrum features.
To achieve this, we improved the initial fitting method by simultaneously fitting the $I$ and $V$ spectra with additional Gaussian components, which significantly reduced the residuals present in previous approaches.
For model selection, we employed the Akaike information criterion~\citep[AIC,][]{aic} and Bayesian information criterion~\citep[BIC,][]{bic} to balance the goodness-of-fit against the model complexity. The AIC is calculated as $Nln(\sum(y_i^{model}-y_i^{obs})^2/N)+2k$, and the BIC $=Nln(\sum(y_i^{model}-y_i^{obs})^2/N)+kln(N)$, where $N$ represents the number of spectral channels, $k$ is the number of free parameters, and $y_i^{model}$ and $y_i^{obs}$ denote the modeled and observed flux densities in the $i$th channel respectively. These criteria enabled us to fit spectral features above 3$\sigma_T$ significance (see Fig.~\ref{gaufit}) while minimizing the number of Gaussian components required to achieve reasonable residuals.

The lower panel of Fig.~\ref{gaufit} shows that the final model incorporates 25 Gaussian components and successfully reproduces the $I$ and $V$ spectra. The corresponding parameters are listed in Table~\ref{tabfit2}. To estimate the parameter uncertainties, we employed a bootstrap resampling approach by separating the {\it IRAS}\,02524+2046 observational data into 110 files, each of which contained the observational results of one minute on-source integration. 
We generated 1\,000 bootstrap samples for the $I$ and $V$ spectra by random resampling. The fitting results of the 1\,000 samples were used to calculate parameter errors.

Compared to the first method, our analysis reveals two additional Gaussian components in the velocity range $v_{helio}\sim$~54\,050 $-$ 54\,250~km~s$^{-1}$, corresponding to complex features in the $I$ and $V$ profiles. 
We detect significant magnetic fields ($>$ 3$\sigma$) for ten components: eight components from the 1667~MHz transition, and two components from the assumed 1665~MHz line as given in Table~\ref{tabfit2}. The measured field strengths range from $-$24.5~mG to +20.6~mG, with a predominance of positive values (8/10 cases).
To verify these results, we independently analyzed data from both observational epochs and found consistent magnetic field measurements within 3$\sigma$ uncertainties, as tabulated in Table~\ref{tabfit2}. 
For four of the five components with previously reported confident detections by \citet[][their Gaussians 3, 8, 12, and 13]{mh13}, our measurements agree within 3$\sigma$ uncertainties. However, we obtained negative field values for the remaining one component, where \citet[][]{mh13} reported positive fields.
Based on earlier findings, the emissions in the velocity range $v_{\rm helio} \sim$~54\,400~$-$~54\,725~km~s$^{-1}$ likely represent a blend of the 1665~MHz and 1667~MHz OH lines. We therefore reapplied the fitting procedure by adopting a splitting coefficient $b_i = 1.964$ Hz $\mu$G$^{-1}$. With the exception of Gaussian components 4 and 5, the derived magnetic fields are consistent with the values reported in Table~\ref{tabfit2}. Specifically, the best-fit solutions for components 4 and 5 yield larger magnetic field strengths and associated uncertainties: 6.6$\pm$1.3~mG and 12.1$\pm$2.8~mG, respectively.

In the fitting results of Stokes $V$ spectra (Fig.~\ref{gaufit}), the narrow emission line near $v_{\rm helio} \sim$~54\,327~km~s$^{-1}$ cannot be well fit by the above methods, warranting further observational investigation. 
The narrow emission line shows a velocity offset of about 0.5~km~s$^{-1}$ between its Stokes $I$ peak ($v_{\rm helio} \sim54\,326.96$~km~s$^{-1}$) and Stokes $V$ peak ($v_{\rm helio}\sim54\,326.42$~km~s$^{-1}$), with a line width of approximately 2.6~km~s$^{-1}$ in the Stokes $I$ spectrum.
One possible explanation is that it corresponds to one component of a Zeeman pair whose LCP and RCP components are split by an interval larger than the line width, and only the stronger component is detected. As observed in Galactic OH masers \citep[e.g.,][]{gcm+15}, the ratios of the peak flux densities between the LCP and RCP components for Zeeman pairs can range from 0.3 to 2.4.

\section{Discussions and conclusions}
\label{sect:con}

We developed polarization calibration procedures for $L$-band spectral observations and applied them to FAST data of the OH megamaser galaxy {\it IRAS}\,02524+2046. Through complete Mueller matrix solutions, we achieved a circular polarization calibration with an accuracy of $\sim$0.01\%~$-$~0.08\% across the 1050$-$1450~MHz frequency range. We also emphasize that regular calibrator observations are necessary to maintain the polarization calibration accuracy.

The FAST observations of {\it IRAS}\,02524+2046 revealed detailed OH megamaser features in the Stokes $I$ spectrum, including a narrow emission line component with a line width of 2.6~km~s$^{-1}$, two emission line components showing large velocity shifts from the systemic galaxy velocity, and multiple OH emission peaks resolved by the high-spectral resolution.
In addition, a narrow emission line feature near the expected frequency of the redshifted 1720~MHz OH line was detected, making {\it IRAS}\,02524+2046 a new galaxy with a detected OH satellite line.

The detection of the 1665~MHz OH megamaser feature at the heliocentric velocity $\sim$54\,725~km~s$^{-1}$ is particularly significant and suggests that the previously observed unusual flux ratio of the pair of 1665 and 1667~MHz OH megamasers likely results from blended emission of both transitions at similar observed frequencies.
Our analysis revealed that the 1667~MHz OH megamaser emission lines in {\it IRAS}\,02524+2046 spans an exceptionally wide velocity range from $v_{\rm helio}$~$\sim$~54\,750 to $\sim$~53\,580~km~s$^{-1}$, indicating greater complexity than previously recognized. These observations imply that some maser-emitting clumps exhibit large velocity offsets from the systemic galaxy velocity, which is intriguing for OH megamaser galaxies \citep[e.g.,][]{hag16}.
Possible explanations include outflows driven by active galactic nuclei, nuclear starbursts, combined effects of active galactic nuclei and starbursts \citep[e.g.,][]{gfs17}, an association with dual galactic nuclei, or a molecular ring orbiting a supermassive black hole \citep[e.g.,][]{hag16}.

The sensitive polarization observations by FAST reveal detailed local features in the Stokes $V$ profiles of the 1665 and 1667~MHz OH megamasers from {\it IRAS}\,02524+2046, including distinct peaks and dips. 
We simultaneously fit the $I$ and $V$ spectra to decompose the OH megamaser emissions into multiple Gaussian components, identifying ten components with significant Zeeman splitting ($> 3\sigma$). The derived magnetic field strengths span $-$24.5~mG to +20.6~mG, with eight components showing positive values.

The case study of {\it IRAS}\,02524+2046 demonstrated that sensitive polarization observations with a high spectral resolution are essential for resolving individual maser components and their Zeeman splitting features.
These results also indicate that magnetic field measurements derived from some of the individual Gaussian components in single-dish observations require careful interpretation. The overall preferential magnetic field orientation of a galaxy likely provides more reliable physical insight, although high angular resolution very long baseline observations may ultimately be needed to reliably measure the characteristics of the magnetic fields traced by OH megamasers.

\begin{acknowledgements}
We thank the anonymous referee for the very careful reading and helpful suggestions.
This work is supported by the National SKA Program of China (Grant No. 2022SKA0120103) and the National Natural Science Foundation
  of China (Grant No. 12588202 and 11933011). XYG and JLH are
  additionally supported by the International Partnership Program of
  Chinese Academy of Sciences, Grant No. 114A11KYSB20170044. TH thanks the support from the Youth Innovation Promotion Association CAS. LGH
  thanks for the helpful discussions with Dr. W.C. Jing for the
  polarization calibration, and Dr. J. Xu for correcting the influence of the Earth's ionosphere on the calibration results. This work made use of the data from FAST (Five-hundred-meter Aperture Spherical radio Telescope)(https://cstr.cn/31116.02.FAST). FAST is a Chinese national mega-science facility, operated by the National Astronomical Observatories, Chinese Academy of Sciences.

\end{acknowledgements}

\bibliographystyle{aa}
\bibliography{ref}

\begin{appendix} 

\section {Parameters of Mueller matrix for the central beam}
\label{muellerpa}

\begin{figure}[b]
  \centering
  \includegraphics[width=0.49\textwidth]{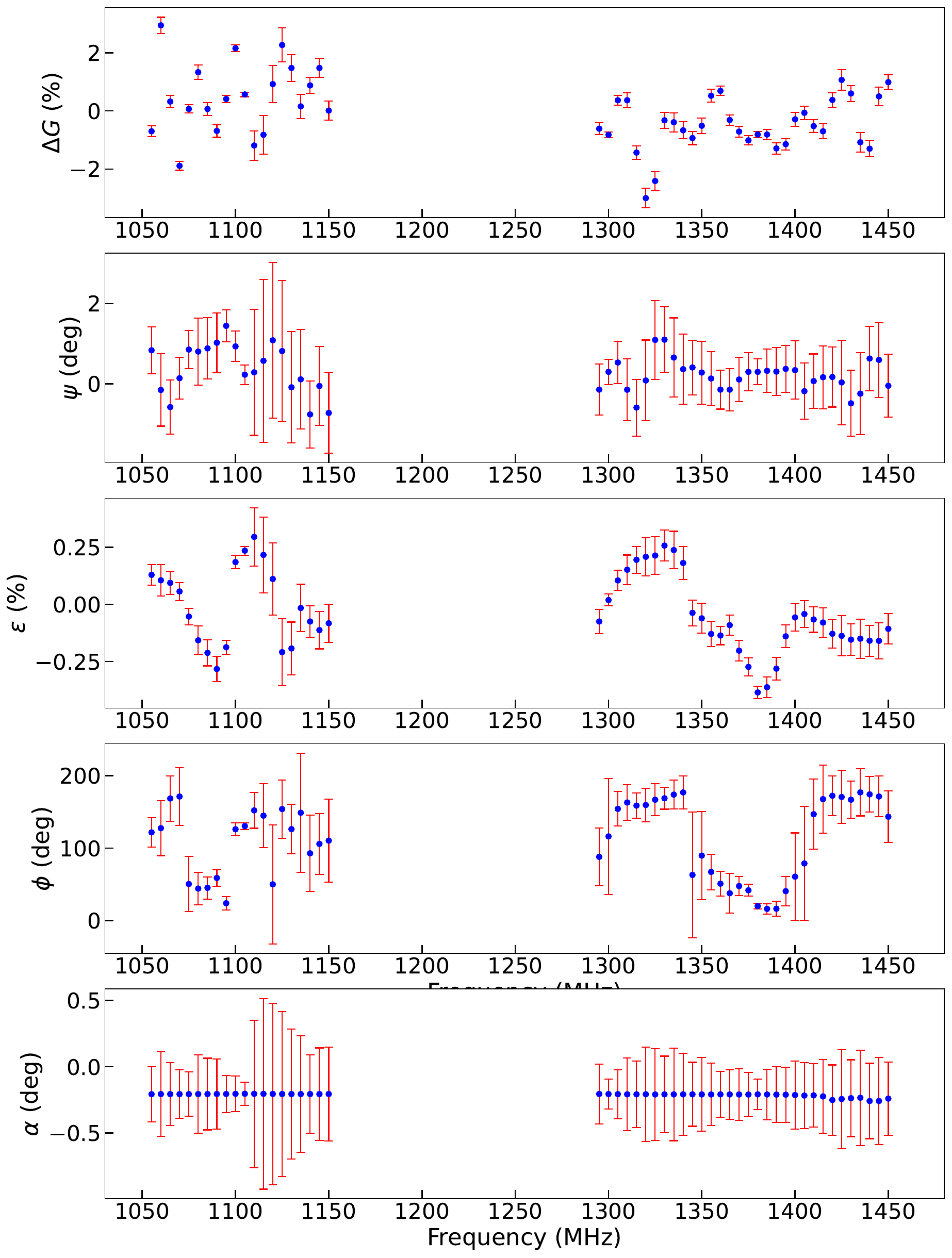} \\
   \caption{Frequency dependence of the five fitted Mueller matrix parameters ($\Delta G$, $\psi$, $\epsilon$, $\phi$, and $\alpha$) across the 1050$-$1150 MHz and 1300$-$1450 MHz bands. Data are derived from FAST observations of 3C\,286 on 19 August 2023.}
   \label{pamuller}
\end{figure}

The fitted values and their associated uncertainties for the five Mueller matrix parameters ($\Delta G$, $\psi$, $\epsilon$, $\phi$, and $\alpha$) are summarized in the corresponding panels in Fig.~\ref{pamuller} for the 1050$-$1150~MHz and 1300$-$1450~MHz frequency bands.
As discussed in \citet[][]{chl+24}, the parallactic angle $\theta$ and the parameter $\alpha$ are coupled in the solution of the Mueller matrix. A small systematic error in $\theta$ may propagate to the fitted $\alpha$ values. In the lower panel of Fig.~\ref{pamuller}, the nearly constant fitted $\alpha$ values across different frequencies likely indicate a systematic error in $\theta$ originating from imperfections in the mechanical control of the receiver rotation.

\section {Baseline and standing wave removal in spectral data processing}
\label{swfitting}

\begin{figure*}[t]
  \centering
  \includegraphics[width=0.93\textwidth]{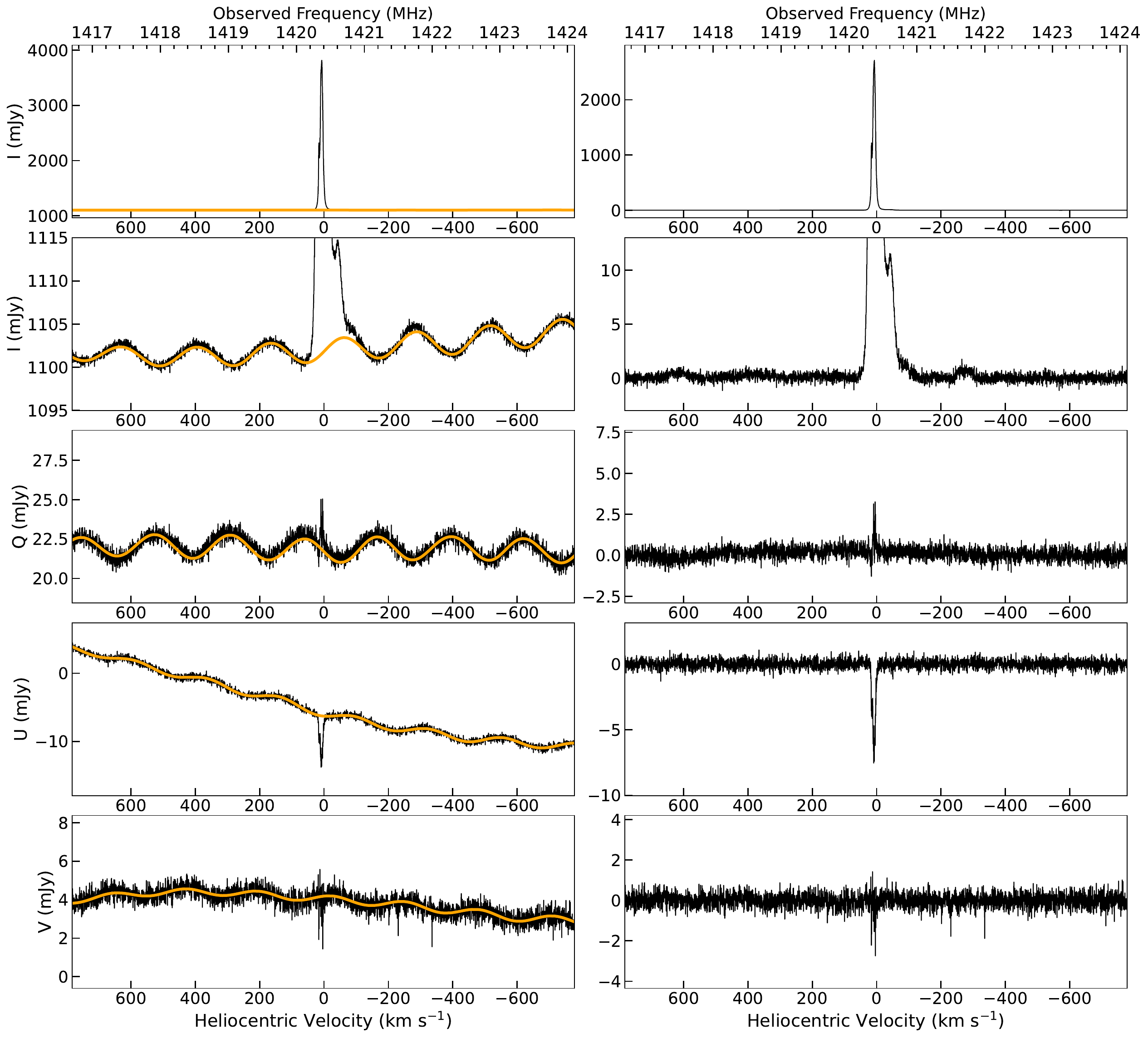} \\
   \caption{{\it Left panels} show the Stokes $I$, $Q$, $U$, and $V$ spectra (black) of Galactic $\hi$ 21~cm emission toward {\it IRAS} 02524+2046 ($l=158.0^\circ$, $b=-$33.3$^\circ$) observed by FAST on 12 August 2023, with orange curves indicating the fitted baseline and standing wave models. The {\it right panels} present the spectra after subtracting these components.}
   \label{himuller}
\end{figure*}

\begin{figure*}
  \centering
  \includegraphics[width=0.93\textwidth]{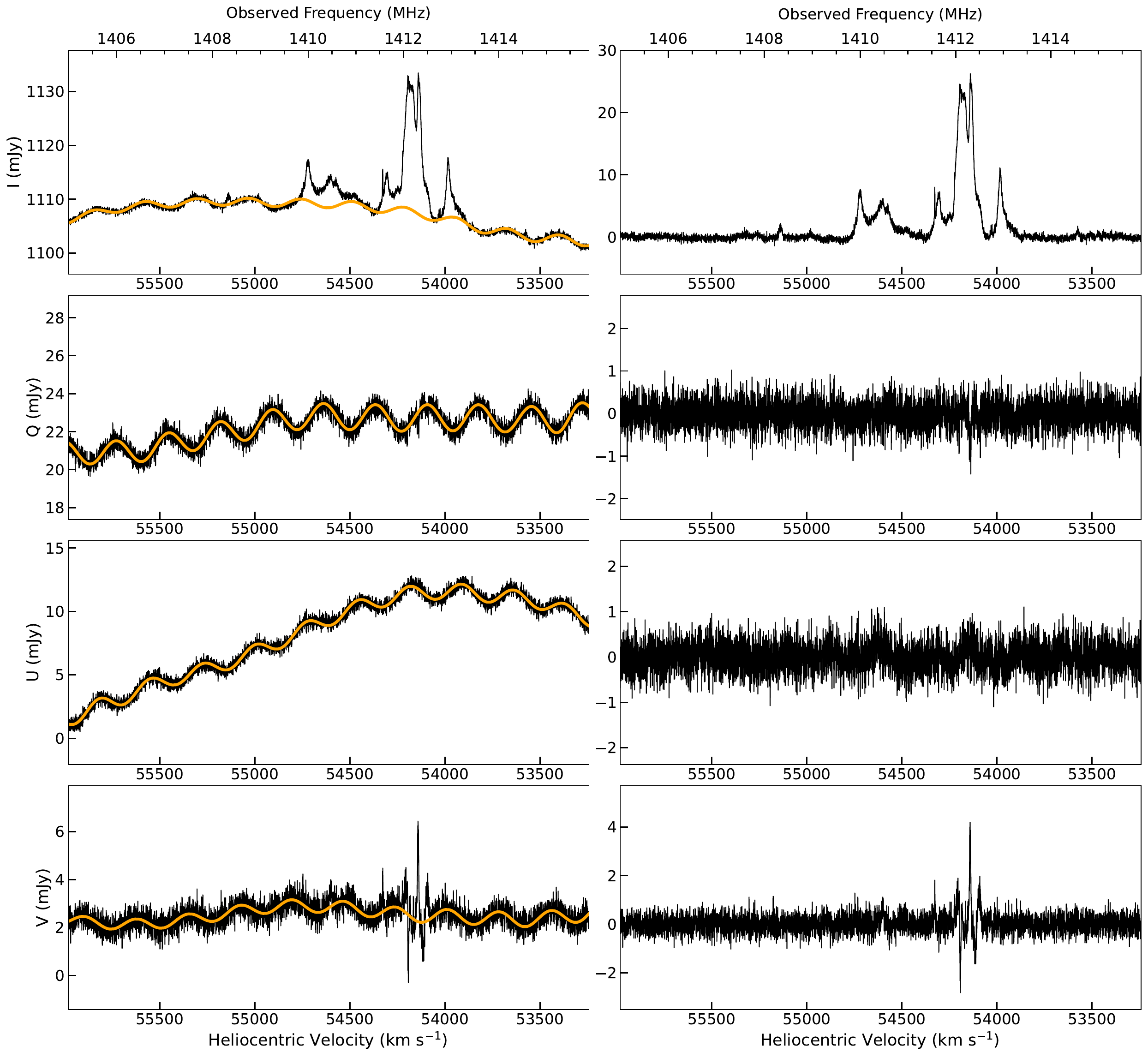} \\
   \caption{Same format as Fig.~\ref{himuller}, showing the results for the 1665~MHz and 1667~MHz OH megamaser emission from {\it IRAS} 02524+2046 observed by FAST on 12 August 2023.}
   \label{ohmuller}
\end{figure*}

We employ a modified sinusoidal function \citep[][]{jhh+23} combined with third-order polynomials to model and remove standing waves and spectral baselines. Fig.~\ref{himuller} and ~\ref{ohmuller} demonstrate this technique through two applications: (1) Galactic $\hi$ 21~cm lines toward {\it IRAS}\,02524+2046 (Galactic coordinates $l=158.0^\circ$, $b=-$33.3$^\circ$), and (2) OH megamasers in {\it IRAS}\,02524+2046.

\end{appendix}

\end{document}